\documentclass[]{IEEEtran}

\usepackage[utf8]{inputenc} 
\usepackage[T1]{fontenc}
\usepackage{url}              
\usepackage{ifthen}          

\usepackage[cmex10]{amsmath}  
\usepackage{graphicx}         
\usepackage{booktabs}         

\usepackage{algpseudocode}    
\algrenewcommand\algorithmicindent{0.5em}

\usepackage[caption=false,font=footnotesize]{subfig}

\usepackage{textcomp}
\usepackage{cite}
\usepackage{mathtools}
\graphicspath{ {./imgs/} }
\usepackage{amssymb,latexsym,amsmath}
\allowdisplaybreaks
\usepackage{enumitem}
\usepackage{float}

\usepackage{xcolor}
\usepackage{verbatim}
\usepackage{tikz}
\usetikzlibrary{shapes.geometric, arrows}
\tikzstyle{arrow} = [thick,->,>=stealth]

\newtheorem{theorem}{Theorem}
\newtheorem{corollary}{Corollary}
\newtheorem{proposition}{Proposition}
\newtheorem{lemma}{Lemma}
\newtheorem{definition}{Definition}
\newtheorem{assumption}{Assumption}

\allowdisplaybreaks

\DeclareMathOperator*{\argmin}{argmin}

\def\BibTeX{{\rm B\kern-.05em{\sc i\kern-.025em b}\kern-.08em
    T\kern-.1667em\lower.7ex\hbox{E}\kern-.125emX}}

\begin{document}
\title{Reinforcement Learning for  Near-Optimal Design of Zero-Delay Codes for \\ Markov Sources}

\author{Liam Cregg, Tam\'as~Linder,~\IEEEmembership{Fellow,~IEEE,}
  and~Serdar~Y\"uksel,~\IEEEmembership{Senior Member,~IEEE}%
    \thanks{This work was supported in part by the Natural Sciences and Engineering Research Council of Canada. The first author was supported by a Queen's University Department of Mathematics and Statistics Summer Research Award.}%
    \thanks{The authors are with the Department of Mathematics and Statistics, Queen's University, Kingston, ON K7L 3N6, Canada (e-mail: liam.cregg@queensu.ca; tamas.linder@queensu.ca; yuksel@queensu.ca).}}

\maketitle

\begin{abstract}
    In the classical lossy source coding problem, one encodes long
    blocks of source symbols that enables the distortion to approach the
    ultimate Shannon limit. Such a block-coding approach introduces
    large delays, which is undesirable in many delay-sensitive
    applications. We consider the zero-delay case, where the goal is to
    encode and decode a finite-alphabet Markov source without any delay. It has been
    shown that this problem lends itself to stochastic control
    techniques, which lead to existence, structural, and general
    structural approximation results. However, these techniques so far
    have only resulted in computationally prohibitive algorithmic
    implementations for code design. To address this problem, we
    present a practically implementable  reinforcement
    learning design algorithm and rigorously prove its asymptotic
    optimality. In particular, we show that a quantized Q-learning
    algorithm can be used to obtain a near-optimal coding policy for
    this problem. The proof builds on recent results on quantized
    Q-learning for weakly Feller controlled Markov chains whose
    application necessitates the development of supporting technical
    results on regularity and stability properties, and relating the
    optimal solutions for discounted and average cost infinite horizon
    criteria problems. These theoretical results are supported by
    simulations.
  \end{abstract}

\section{Introduction}\label{section:introduction}
\subsection{Zero-Delay Lossy Coding}

We consider the problem of encoding an information source without
delay, sending the encoded source over a discrete noiseless channel,
and reconstructing the source, also without delay, at the
decoder. Hence, the classical block-coding approach is not allowed. Zero-delay coding schemes
have many practical applications in emerging fields such as
networked control systems (see \cite{YukselBasarBook} and references therein for
an extensive review and discussion of applications), real-time
mobile audio-video systems (as in streaming systems \cite{Draper}),
and real-time sensor networks \cite{Akyildiz}, among other areas.

\subsubsection*{Notation}

\begin{itemize}
  
\item[] We will use the subscript $t$ to denote a time-dependent object
  (as the samples $X_t$ of an information source); any other
  subscript denotes some other dependency. In particular, the subscript
  $n$ will be used to denote an underlying quantization parameter, and
  should not be confused with time.  Probabilities and expectations
  will be denoted by $P$ and $\mathbf{E}$, respectively. When the relevant distributions depend on some parameters, we include these in the superscript and/or subscript.

\item[] Random variables will in general be denoted by uppercase
  letters. We make a few exceptions in order to conform with the
  corresponding literature on zero-delay coding; in particular, the
  (random) sequence of quantizers and channel symbols will be denoted by $Q_t$
  and $q_t$ respectively. However, they can be distinguished from
  their realizations by the time subscript. For example, we write $P(Q_t = Q)$
  and $P(q_t = q)$. When discussing (time-homogeneous) Markov processes, we will
  often use the shorthand $P(x'|x) = P(X_{t+1} = x' | X_t = x)$. In
  the case of multiple Markov chains, which transition probability we
  mean will be clear from the variable names. For example, if we have
  two Markov chains $\{X_t\}_{t \ge 0}$ and $\{Y_t\}_{t \ge 0}$, then
  $P(x'|x)$ and $P(y'|y)$ are the respective transition probabilities.
      
\item[] We use superscripts to denote the product sets,
      (e.g.\  $\mathbb{X}^t$ denotes the $t$-fold product of the
      source alphabet $\mathbb{X}$), and for random vectors taking values in
      these sets we use the notation
      $X_{[0,t-1]} \coloneqq (X_0, \ldots, X_{t-1})$. Also, we use
      $\mathcal{P}(\mathbb{X})$ to denote the space of probability
      measures over the set $\mathbb{X}$ (and we endow this space with
      the topology of weak convergence).
\end{itemize}

The source $ \{X_t\}_{t \ge 0} $ is a time-homogeneous, discrete-time
Markov process taking values in a finite set $\mathbb{X}$ and has
transition matrix $ P(x' | x) $. We assume that the source is
irreducible and aperiodic, and thus admits a unique invariant measure,
which we will denote by $\zeta$. We also assume that the distribution
of $X_0$, which we denote by $\pi_0$ (this can be different from
$\zeta$), is available at the encoder and decoder.

At time $t\ge 0$, the encoded (compressed) information, denoted by
$q_t$, is sent over a discrete noiseless channel with common input and
output alphabet $ \mathcal{M} \coloneqq \{1,\ldots,M\} $. The encoder
is defined by an encoder policy
$ \gamma^e = \{\gamma^e_t\}_{t \ge 0} $, where
$ \gamma^e_t : \mathcal{M}^t \times \mathbb{X}^{t+1} \to \mathcal{M}
$, and $q_t = \gamma^e_t(q_{[0,t-1]},X_{[0,t]})$. Note that given $q_{[0,t-1]}$ and $X_{[0,t-1]}$,
the map $\gamma^e_t(q_{[0,t-1]}, X_{[0,t-1]},\, \cdot\, ) $ is a
\emph{quantizer} (i.e.\ a map from $\mathbb{X}$ to $\mathcal{M}$),
which we denote by $Q_t$. We will denote the set of all quantizers
from $\mathbb{X}$ to $\mathcal{M}$ by
$\mathcal{Q}$. Thus we can view an encoder policy at time $t$ as
selecting a quantizer $Q_t \in \mathcal{Q}$ and then encoding
(quantizing) $X_t$ as $q_t = Q_t(X_t)$. We call such encoder policies
admissible, and denote the set of all admissible encoder policies
(sometimes called quantization policies) by $\Gamma^e$. Upon receiving $q_t$, the decoder generates the reconstruction $ \hat{X}_t $ without
delay, using decoder policy $ \gamma^d = \{\gamma^d_t\}_{t \ge 0} $,
where $ \gamma^d_t : \mathcal{M}^{t+1} \to \hat{\mathbb{X}} $ and
where $\hat{\mathbb{X}}$ is a finite reproduction alphabet. Thus
we have $ \hat{X}_t = \gamma^d_t(q_{[0,t]}) $. 
The set of these admissible decoder policies is denoted by~$\Gamma^d$.

In general for the zero-delay coding problem, the goal is to minimize the average distortion (cost), given by
\begin{align}\label{eq:average_cost}
    J(\pi_0, \gamma^e, \gamma^d) \coloneqq
    \limsup_{T\to\infty}\mathbf{E}_{\pi_0}^{\gamma^e, \gamma^d} \left[ \frac{1}{T}\sum_{t=0}^{T-1}d(X_t,\hat{X}_t)\right],
\end{align}
where $d : \mathbb{X} \times \hat{\mathbb{X}} \to [0, \infty)$ is a
given distortion measure and $\mathbf{E}_{\pi_0}^{\gamma^e, \gamma^d}$ is the
expectation with initial distribution $X_0 \sim \pi_0$ under encoder
policy $\gamma^e$ and decoder policy $\gamma^d$.

It is straightforward to show that, for a fixed encoder policy
$\gamma^e$, the optimal decoder policy for all $t\ge 0$ is given by
\begin{equation}
    \gamma^{d*}_t(q_{[0,t]}) = \argmin_{\hat{x} \in \hat{\mathbb{X}}}\mathbf{E}^{\gamma^e}_{\pi_0}\left[ d(X_t,\hat{x}) | q_{[0,t]} \right].\label{eq:decoder}
\end{equation}

Thus we identify a coding
policy with the corresponding encoder policy by assuming that an
optimal decoding policy is used for any given encoding policy  and
will denote $ \gamma \coloneqq \gamma^e $ and
$ \Gamma \coloneqq \Gamma^e$. We can then restrict our search to
finding optimal encoding policies. With an abuse of notation,  we denote
\begin{align*}
    J(\pi_0, \gamma) \coloneqq \inf_{\gamma^d \in \Gamma^d} J(\pi_0, \gamma^e, \gamma^d).
\end{align*}
The objective is then to minimize $J(\pi_0, \gamma)$ over all
$\Gamma$. We will denote the optimal cost by
\[
    J^*(\pi_0) \coloneqq
    \inf_{\gamma \in \Gamma}J(\pi_0, \gamma).
\]

We will also consider the discounted cost problem, which is the minimization of
\begin{align}\label{eq:discounted_cost}
    J_{\beta}(\pi_0, \gamma^e, \gamma^d) \coloneqq
    \lim_{T\to\infty}\mathbf{E}_{\pi_0}^{\gamma^e, \gamma^d} \left[\sum_{t=0}^{T-1}\beta^t d(X_t,\hat{X}_t)\right].
\end{align}
for some $\beta \in (0,1)$. As above, we assume an optimal decoder
policy and minimize only over the encoder policies, yielding
$J_\beta(\pi_0, \gamma)$ and
$J_\beta^*(\pi_0) \coloneqq \inf_{\gamma \in \Gamma}J(\pi_0,
\gamma)$. We note that, as opposed to the optimal average distortion
$ J^*(\pi_0)$, the quantity $J_\beta^*(\pi_0)$ has little importance
from a source coding point of view and we only use it as a tool toward
designing codes that are near-optimal in the average distortion
sense.

We say that a set of policies $\{\gamma\}$ depending on some parameter
set is  \emph{near-optimal} if  for any
$\epsilon > 0$, there is some choice of parameters  (to be identified
explicitly later) such that the resulting policy  $\gamma$ satisfies
$J(\pi_0, \gamma) \le J^*(\pi_0) + \epsilon$.

\subsection{Literature Review}
A number of important results have been obtained in the literature concerning  the
structure of optimal zero-delay codes, starting with the foundational
papers by Witsenhausen~\cite{Witsenhausen} and Walrand and
Varaiya~\cite{WalrandVaraiyaControl}. In particular, for the finite
horizon problem,~\cite{Witsenhausen} showed that any encoder policy
can be replaced, without performance loss, by one using only
$q_{[0,t-1]}$ and $X_t$ to generate
$q_t$. Furthermore,~\cite{WalrandVaraiyaControl} proved a similar
result for an encoder policy using only the conditional probability
$P(X_t \in \, \cdot\, | q_{[0,t-1]})$ and $X_t$ to generate
$q_t$. These results have been further generalized, see
e.g., \cite{Teneketzis, MahTen09, YukIT2010arXiv, YukLinZeroDelay,
  wood2016optimal}. In particular,~\cite{wood2016optimal} showed the
existence of optimal policies in the infinite-horizon case, which we
now review.

Recall that $\mathcal{P}(\mathbb{X})$ denotes the space of probability measures over $\mathbb{X}$, and define the conditional probability $\pi_t \in \mathcal{P}(\mathbb{X})$ as the ``belief'' on $X_t$ given $q_{[0,t-1]}$, i.e.
\begin{equation}\label{eq:condProbM}
    \pi_t(A) \coloneqq P^{\gamma}_{\pi_0}(X_t \in A | q_{[0,t-1]})
\end{equation}
for all measurable $A\subset \mathbb{X}$. Define $\overline{\pi}_t$ similarly, but conditioned on $q_{[0.t]}$, i.e.
\begin{equation}
    \overline{\pi}_t(A) \coloneqq P^\gamma_{\pi_0}(X_t \in A | q_{[0,t]}).\label{eq:filter}
\end{equation}
Note that $\pi_t$ and $\overline{\pi}_t$ are often called the \emph{predictor} and the \emph{filter}, respectively.

\begin{definition}
    We say an encoder policy $\gamma = \{\gamma_t\}_{t\ge0}$ is of the \emph{Walrand-Varaiya type} if, at time $t$, the policy uses only $\pi_t$ and $X_t$ to generate $q_t$. That is, $\gamma$ selects a quantizer $Q_t = \gamma_t(\pi_t)$ and $q_t$ is generated as $q_t = Q_t(X_t)$. Such a policy is called \emph{stationary} if it does not depend on $t$. We denote the set of stationary Walrand-Varaiya type policies by $\Gamma_{\text{WS}}$.
\end{definition}

Note that $\pi_t$ and $\overline{\pi}_t$ can be obtained recursively from
$\pi_{t-1}$, $q_{t-1}$, and $Q_{t-1}$ (see the update equation
\eqref{eq:pi_update}, see also \cite{Chigansky, Handel}). Since the initial distribution $\pi_0$ is known
to both the encoder and decoder, the decoder can track  $\pi_t$, $\overline{\pi}_t$ and
$Q_t$ for all $t\ge 0$. The overall coding scheme for a stationary Walrand-Varaiya encoder is summarized in Figure~\ref{fig:diagram}. Recall that the optimal decoder is given by \eqref{eq:decoder}, and note that the expectation in \eqref{eq:decoder} is simply an expectation with respect to $\overline{\pi}_t$.

\tikzstyle{block} = [rectangle, rounded corners, minimum width=.5cm, minimum height=.5cm,align=center, draw=black]
\tikzstyle{openblock} = [minimum width=.5cm, minimum height=.5cm,align=center]

\begin{figure}[h!]
\centering
\begin{tikzpicture}[node distance=2cm]
    \node (source) [openblock] {$X_t$};
    \node (encoder) [block, right of=source, xshift=1.5cm] {\underline{Encoder}\\$Q_t = \gamma(\pi_t)$ \\ $q_t = Q_t(X_t)$};
    \node (decoder) [block, below of=encoder] {\underline{Decoder}\\$\hat{X}_t = \argmin_{\hat{x}}\sum_x d(x,\hat{x})\overline{\pi}_t(x)$};
    \node (reconstruction) [openblock, left of=decoder, xshift=-1.5cm] {$\hat{X}_t$};
    \node (q) [openblock, below of=encoder, xshift=3.3cm, yshift=.9cm] {$q_t$};
    \draw [arrow] (source) -- (encoder);
    \draw [arrow] (encoder) -- +(3,0) |- (decoder);
    \draw [arrow] (decoder) -- (reconstruction);
\end{tikzpicture}
\caption{Block diagram for our zero-delay coding system: $X_t$ is the
  source sample, $q_t$ is the encoded  symbol transmitted through the
  noiseless channel, and $\hat{X}_t$ is the reconstructed source sample.}
\label{fig:diagram}
\end{figure}
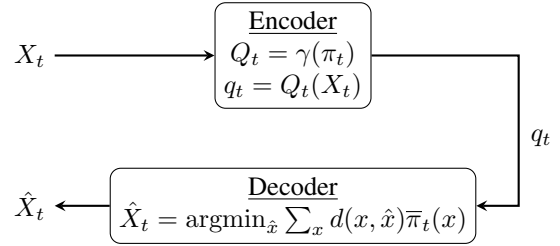

An important fact regarding the Walrand-Varaiya type formulation of
zero-delay coding of Markov sources is that the belief process
$\{\pi_t\}_{t\ge 0}$ can be considered as a $\mathcal{P}(\mathbb{X})$-valued
Markov Decision Process (MDP) with
$\mathcal{Q}$-valued control $\{Q_t\}_{t\ge 0}$, see, e.g., \cite{YukIT2010arXiv}
or \cite{wood2016optimal} for a proof. This observation was
fundamental in deriving the  following results. 

\begin{proposition}\label{prop:Wood}\cite[Theorem 3]{wood2016optimal}
    There exists an optimal policy $\gamma^* \in \Gamma_{\text{WS}}$ for the average cost problem (\ref{eq:average_cost}).  That is, there exists $\gamma^*\in
        \Gamma_{\text{WS}}$ such that 
    \[J(\pi_0,\gamma^*) = J^*(\pi_0)   \text{\ \ for all $\pi_0$}.     \]
\end{proposition}

\begin{proposition}\label{prop:Wood2}\cite[Proposition 2]{wood2016optimal}
    For any $\beta\in (0,1)$, there exists an optimal policy $\gamma_\beta^* \in
        \Gamma_{\text{WS}}$ for the discounted cost problem
    (\ref{eq:discounted_cost}). That is, there exists $\gamma_\beta^*\in
        \Gamma_{\text{WS}}$ such that
    \[J_\beta(\pi_0,\gamma_\beta^*) = J_\beta^*(\pi_0)  \text{\ \ for all $\pi_0$}.
    \]
\end{proposition}

Despite the key structural results 
reviewed above, finding an optimal policy for either finite or infinite
horizons is difficult. For some very special cases, the optimal
solution is known. For example,~\cite{WalrandVaraiyaControl} showed memoryless
encoding is optimal when $\mathbb{X} = \mathcal{M}$ and the channel
is noisy and symmetric. However, for a general source and channel (or
for a noiseless channel, as in this paper), finding an optimal
encoding policy is an open problem.

Stochastic control\footnote{We refer
    the reader, e.g.,  to the texts  
    \cite{Bertsekas}, \cite{HernandezLermaMCP}, and   \cite{Szepesvari} for
    an  introduction into the theory of stochastic control.} based  approaches 
 play an important role in the above  structural results. Under a stochastic control
framework,~\cite{Witsenhausen, WalrandVaraiyaControl} used dynamic
programming,~\cite{wood2016optimal, TatikondaIT, ghomi2021zero} made
use of the value iteration algorithm and the vanishing discount
method, whereas~\cite{YukLinZeroDelay} used the convex analytic method
to obtain structural results for optimal codes. A key component in these
results, and one that we will use in showing the near-optimality of our algorithm, is that the zero-delay coding problem can be restated as
a Markov Decision Process (MDP) with $\{\pi_t\}_{t \ge 0}$ as the
state process. However, there are limitations when using these methods
to obtain an optimal solution. In particular, dynamic programming
relies on backwards induction from a finite time horizon, which is not
applicable for the infinite horizon case. Furthermore, the
implementation of the value iteration algorithm requires the
computation of certain value functions and conditional expectations,
which is practically very challenging due to the probability measure-valued
state dynamics.

These challenges, which will be made more explicit in
Section~\ref{section:optimal quantizers},  motivate the use of a
reinforcement learning approach that we present in this paper. A
popular reinforcement learning algorithm,
Q-learning~\cite{Watkins,TsitsiklisQLearning,Baker,CsabaLittman,Szepesvari,BertsekasTsitsiklisNeuro}
is primarily used for fully observed finite space MDPs. This algorithm
does not require the knowledge of the transition kernel, or even
the cost (or reward) function, for its implementation. In this
algorithm, the incurred per-stage cost variable is observed through the
simulation of a single sample path. When the state and action spaces
are finite, under mild conditions that require that all state-action
pairs are visited infinitely often, this algorithm is known to
converge to the optimal cost. Recently, this algorithm has been
generalized to be applicable for continuous space MDPs
(see~\cite{KSYContQLearning} and the references therein).

In the broader literature related to zero-delay coding, often
information theoretic relaxation techniques are used to convexify the
non-convex zero-delay optimal quantization problem, which lead to
lower bounds on optimal performance, as well as to upper bounds. These
include replacing the number of bins constraint with a mutual
information constraint, applying the Shannon lower bounding technique,
or entropy coding (see, e.g., \cite{Silva1},
\cite{silva2015characterization} \cite{stavrou2018zero}). Using
ergodicity and invariance properties, \cite{cuvelier2022time} has
constructed time-invariant coding schemes using dithering. A further
line of work for linear systems follow the sequential rate-distortion
theoretic approach
\cite{banbas89,TatikondaSahaiMitter, tanaka2016semidefinite,
  stavrou2018zero,derpich2012improved,stavrou2021asymptotic}. For
coding of Gaussian sources over additive Gaussian channels, some of
these results become operational for zero-delay coding
\cite{banbas89}; see also
\cite{tanaka2016semidefinite},\cite{stavroutanaka2019time}, and
\cite{kostina2019rate}.

We note that applying learning theoretic methods in the theory of
optimal (lossy) source coding has prior history. A well established
line of study of this problem focuses on empirical learning methods
for data
compression~\cite{Pollard,linder1994rates,linder2002learning}, though
often limited to i.i.d source models. Our approach here is
complementary, since we consider a highly structured coding problem
instead of (unstructured) vector quantization and we consider sources
with memory. Our analysis leads
to near-optimal solutions directly (without learning the source
distribution). We also refer  to recent  research activity  in
machine learning methods in communications theory (see
e.g.~\cite{gunduz2019machine}); however, our analysis seems to be the
first contribution to source coding in the context of reinforcement learning.

Recently, the approach we introduce and analyze in this paper has been
extended to noisy channels with feedback in \cite{CrAlYu23}. In
\cite{CrAlYu24} an alternative sliding finite window code has been
introduced. Compared with the sliding finite window approach of that
work, the approach in this paper has the following advantages: (i)
quantizing probability measures allows for more relaxed filter
stability conditions (e.g., with no geometric filter stability
conditions); (ii) it avoids the use of a transient period until the
initial memory data is collected. Thus, our approach is complementary
and is 
applicable for a wider class of models. This comes at the cost of
higher computational load due to the Bayesian updates at the encoder during
the implementation prior to the quantization of probability measures.

Finally, stochastic control and reinforcement learning techniques have also been applied to the ``dual'' problem of channel coding. For example, \cite{TatikondaIT, PermuterTrapdoor, PermuterIsing} all use an MDP formulation along with dynamic programming to obtain results on channel capacity and capacity-achieving codes. Learning-theoretic results to this end include \cite{Aharoni, TsurDataDriven}.

\subsection{Contributions}

We formulate the zero-delay coding problem so that it is amenable to a
reinforcement learning approach. In particular, the MDP associated
with our zero-delay lossy coding problem has an uncountable state
space (the set of beliefs) and thus has to be discretized (quantized)
to apply Q-learning. After posing the problem as an MDP, we
build on recent results from~\cite{KSYContQLearning} to rigorously
justify the convergence of a reinforcement learning algorithm to a
near-optimal solution (depending on the discretization on the state
space), first for the discounted cost problem, and then for the
average cost problem. In particular, \cite{KSYContQLearning} showed
that, under mild assumptions, a Q-learning algorithm in which the
state is quantized converges to the optimal solution as the maximum
diameter of the quantization bins for the state space goes to
zero. However, the results of \cite{KSYContQLearning} cannot be
straightforwardly applied to our zero-delay coding problem and there are several additional
ingredients needed for our analysis: (a) The convergence and
near-optimality was shown in \cite{KSYContQLearning} only for the discounted cost
criterion problem. Our focus is the  average cost setup;
this will be addressed via relating discounted cost optimal coding
policies 
to ones that are near-optimal for the average cost. (b) We also need to
prove several technical results that are necessary for applying the algorithm in
\cite{KSYContQLearning}, such as the unique ergodicity under an
exploration policy which has not been studied for our setup. Specifically, our main contributions are the following: 

\begin{itemize}
    \item We present a reinforcement learning algorithm for the
          near-optimal design of stationary zero-delay codes (Algorithm~1). As
          an auxiliary result, we state the near-optimality of the algorithm
          for the discounted cost problem when the source starts from the 
          invariant distribution 
          (Theorem~\ref{theorem:near-optimal}). Then we show that an optimal
          policy for the discounted cost problem (for sufficiently large
          discount parameter) can be used to
          obtain a near-optimal policy for the average cost problem
          (Theorem~\ref{theorem:discount-to-average2}). This gives, to our knowledge, the first
          concrete implementation of a provably near-optimal algorithm for the
          zero-delay coding problem.

        \item To show the convergence of our algorithm, we prove additional
          regularity properties of the MDP formulation of the zero-delay coding
          problem. In particular, we show that the process $\{\pi_t\}_{t \ge 0}$
          is stable under the uniform exploration policy, and then deduce unique
          invariance under this same policy, which is necessary for the
          application of \cite{KSYContQLearning}. Furthermore, unlike
          in 
          \cite{KSYContQLearning}, due to the lack of strong
          recurrence conditions of the predictor process in our setup,
          additional analysis of the initialization is
          necessary, which we also provide.

    \item Finally, Section~\ref{section:results} provides simulations
          comparing Algorithm~1 with the so called omniscent finite-state
          scalar quantization (O-FSSQ) design algorithm \cite[Chapter
              14]{GershoVectorQuant}, a heuristic, but effective technique for
          designing zero-delay lossy codes, to demonstrate the superior
          performance of Algorithm~1.
\end{itemize}

The rest of the paper is organized as follows: In
Section~\ref{section:Algorithms} we present our reinforcement learning
algorithm and in Theorem~\ref{theorem:near-optimal} and
Theorem~\ref{theorem:discount-to-average2} state the near-optimality
of the resulting stationary policies for the discounted and average
cost problems, respectively, when the source starts from its invariant
distribution. Although proofs of the the main results are given in
Section~\ref{section:proofs}, we first have to introduce necessary
concepts and auxiliary results that will be needed in these proofs. In
particular, in Section~\ref{section:optimal quantizers} we review how
the zero-delay coding problem can be turned into a problem involving
an MDP, and in Section~\ref{section:Q-learningC} we present the
quantized Q-learning algorithm in \cite{KSYContQLearning} in the
context of a general MDP. In Section~\ref{section:unique-ergodicity}
we prove that the zero-delay coding MDP meets the necessary
assumptions to apply the quantized Q-learning algorithm, in particular
the unique ergodicity of our MDP under the uniform exploration
policy. Section~\ref{section:proofs} contains the proofs of the main
results and Section~\ref{section:results} presents simulation
results. Conclusions are drawn in Section~\ref{sec-concl}, where
future research directions are also discussed.  Some technical results
relating discounted cost optimal and average cost optimal policies are
relegated to Appendix \ref{appendix:A}.

\section{Near-Optimal Design of Zero-Delay Codes}\label{section:Algorithms}
In order to make the algorithm self-contained, we first introduce some
definitions and update equations. The rationale for these will be
formalized during the proof of convergence to near-optimality, but we give a high-level justification in this section. Note that the implementation of our algorithm does not require the technical knowledge of MDPs used in the proofs, so we first present our algorithm in its entirely and then prove its near-optimality in the following sections.

Recall the definition of the belief $\pi_t$
in~\eqref{eq:condProbM}. Under a Walrand-Varaiya type policy,
$\pi_{t+1}$ can be obtained from $\pi_t$, $q_t$, and $Q_t$ via the update
equation \cite{YukLinZeroDelay}
\begin{equation}
    \pi_{t+1}(x') = \frac{1}{\pi_t(Q_t^{-1}(q_t))}\smashoperator[r]{\sum_{x \in Q_t^{-1}(q_t)}}P(x'|x)\pi_t(x),\label{eq:pi_update}
\end{equation}
where $Q_t^{-1}(q_t)= \{x\in \mathbb{X}: Q_t(x)=q_t\}$ and
$\pi_t(Q_t^{-1}(q_t)) = \sum_{x \in Q_t^{-1}(q_t)} \pi_t(x)$. Note
that the encoder and decoder can both track $\pi_t$ and thus $Q_t$, so
these quantities are known at time $t$. This update essentially
performs a Bayesian update on $\pi_t$ to compute $\overline{\pi}_t$,
then computes $\pi_{t+1}(x') = \sum_{x}P(x'|x)\overline{\pi}_t(x)$.

Recall that $\mathcal{Q}$ is the set of all quantizers from $\mathbb{X} \to \mathcal{M}$. We wish to compute the ``cost'' of using a given quantizer $Q$ when the predictor is $\pi$. The natural choice is the expected distortion, given $Q$ and $\pi$, using $\gamma^{d*}$ (recall \eqref{eq:decoder}) as the decoder. This yields the following cost function $c : \mathcal{P}(\mathbb{X}) \times \mathcal{Q} \to \mathbb{R}_+$,

\begin{equation}\label{eq:cost}
    c(\pi, Q) \coloneqq \sum_{i=1}^M \min_{\hat{x} \in \hat{\mathbb{X}}} \sum_{x \in Q^{-1}(i)} \pi(x)d(x,\hat{x}).
\end{equation}

We approximate $\mathcal{P}(\mathbb{X})$ with the following finite set. Let $m \coloneqq \left| \mathbb{X} \right|$. Given a fixed parameter $n \in \mathbb{Z}_+$, we define
\begin{align}
    \mathcal{P}_n(\mathbb{X}) \coloneqq \Bigl\{ \hat{\pi} \in
      \mathcal{P}(\mathbb{X}) : & \hat{\pi} = \left( \frac{k_1}{n},
        \ldots, \frac{k_m}{n} \right), \label{eq:Pn} \\
        & k_i \in  \mathbb{Z}_+, \, i=1,\ldots,m \Bigr\}. \nonumber 
\end{align}

Given any $\pi \in \mathcal{P}(\mathbb{X})$, let $\hat{\pi}$ denote
the nearest  neighbour of $\pi$  (in Euclidean distance) in $
    \mathcal{P}_n(\mathbb{X})$. We note that $\hat{\pi}$ can be
effectively calculated  using \cite[Algorithm 1]{Reznik}, which we include as
Algorithm~2 in Appendix \ref{appendix:B} for convenience. This algorithm 
``quantizes'' $\pi\in \mathcal{P}(\mathbb{X})$ to its nearest
neighbor $\hat{\pi} \in \mathcal{P}_n(\mathbb{X})$. 

Finally, consider a
$\mathcal{P}(\mathbb{X}) \times \mathcal{Q}$-valued sequence
$\{\pi_t, Q_t\}_{t \ge 0}$ and the resulting
$\mathcal{P}_n(\mathbb{X}) \times \mathcal{Q}$-valued sequence
$\{\hat{\pi}_t, Q_t\}_{t \ge 0}$, where $\hat{\pi}_t$ is the nearest
neighbor of $\pi_t$ in $\mathcal{P}_n(\mathbb{X})$. The following updates are based on the Q-learning equations in~\cite{KSYContQLearning}. We define $\mathsf{Q}_t$ and $\alpha_t$,
which are both functions from $\mathcal{P}_n(\mathbb{X}) \times \mathcal{Q}$ to
$[0,\infty)$, by
\begin{align}
    \mathsf{Q}_0(\hat{\pi}, Q)        & \equiv 0 \nonumber                                                                                                                                                                                                           \\
    \mathsf{Q}_{t+1}(\hat{\pi}_t,Q_t) & =(1-
                                        \alpha_t(\hat{\pi}_t,Q_t)) \mathsf{Q}_t(\hat{\pi}_t,Q_t) \label{eq:Q-factor-quantized} \\ & +
                                        \alpha_t(\hat{\pi}_t,Q_t)[c(\pi_t,Q_t)
                                        + \beta \; \underset{v \in
                                        \mathcal{Q}}{\text{min}} \;
                                        \mathsf{Q}_t(\hat{\pi}_{t+1},v)] \nonumber 
  \\ 
    \mathsf{Q}_{t+1}(\hat{\pi},Q)     & = \mathsf{Q}_t(\hat{\pi}, Q) \text{ for all } (\hat{\pi}, Q) \neq (\hat{\pi}_t, Q_t)                                                \nonumber                                                                           \\
    \alpha_t(\hat{\pi},Q)             & = \frac{1}{1 + \sum_{k=0}^t
                                        \mathbf{1}((\hat{\pi}_k,Q_k) =
                                        (\hat{\pi},Q))} \label{eq:alpha_update}
\end{align}
where $\mathbf{1}((\hat{\pi}_k,Q_k) = (\hat{\pi},Q))$ is the indicator
function of the pair $(\hat{\pi},Q)$.

With these definitions, we are now ready to introduce our Q-learning algorithm to find a near-optimal encoding policy $\gamma \in \Gamma_{WS}$.


\noindent \underline{\textbf{Algorithm 1: Q-learning for Near-Optimal}} \underline{\textbf{Zero-Delay Quantization}}\label{algorithm:2}

\begin{algorithmic}[1] \label{alg1}
    \Require source alphabet $\mathbb{X}$, channel alphabet
    $\mathcal{M}$, transition kernel $P(x' | x)$, initial
    distribution $\pi_0$, quantization parameter $n$, discount factor
    $\beta\in (0,1)$
    \State Sample $ X_0 $ according to $ \pi_0 $
    \State Quantize $ \pi_0 $ using Algorithm 3 with parameter $n$ to obtain $ \hat{\pi}_0 $
    \State Randomly select quantizer $ Q_0 $ uniformly from $\mathcal{Q}$
    \State $ q_0 = Q_0(X_0) $
    \For{$t\ge0$}
    \State Compute $c(\pi_t, Q_t)$ using~\eqref{eq:cost}
    \State Sample $ X_{t+1} $ according to $ P(x' | x) $
    \State Compute $ \pi_{t+1} $ using~\eqref{eq:pi_update}
    \State Quantize $ \pi_{t+1} $ using Algorithm 3 with parameter $n$ to obtain $ \hat{\pi}_{t+1} $
    \State Update $\mathsf{Q}_{t+1}$ and $\alpha_{t+1}$ using~\eqref{eq:Q-factor-quantized} and~\eqref{eq:alpha_update}
    \State Randomly select quantizer $ Q_{t+1} $ uniformly from $\mathcal{Q}$
    \State $ q_{t+1} = Q_{t+1}(X_{t+1}) $
    \EndFor
\end{algorithmic}

In the above algorithm, the initial distribution $\pi_0$ can be arbitrary.  
Note that the stopping criterion in this algorithm can be any measure
of the convergence of $\mathsf{Q}_t$. In our implementation, we
stop the algorithm  when the pointwise difference between
$\mathsf{Q}_{t+1}$ and $\mathsf{Q}_t$ is below some small threshold,
i.e.
$\max_{\hat{\pi}, Q} \left| \mathsf{Q}_{t+1}(\hat{\pi},Q) -
  \mathsf{Q}_t(\hat{\pi},Q) \right| \le \epsilon$. For results on
convergence time for Q-learning algorithms see
e.g.~\cite{even2004learning}. When choosing $Q_{t+1}$ uniformly from
$\mathcal{Q}$, one could exhaustively compute the set of all
quantizers (this can be done offline) and choose uniformly from
them. An alternative (and equivalent)  method is to randomly
 choose
$q = Q(x)$ from $\mathcal{M}$, according to the uniform distribution,
for each $x \in \mathbb{X}$.

The following two theorems are our main results. Both are proved in
Section~\ref{section:proofs}, but these proofs rely on concepts and
auxiliary results from Sections~\ref{section:Q-learning} and \ref{section:unique-ergodicity}.

The first result shows convergence of Algorithm~1 to a near-optimal
policy for the discounted cost (distortion) problem if the source
starts from the 
the unique invariant  distribution $\zeta$.  Note that in
this case $\{X_t\}_{t \ge 0}$ is a stationary and ergodic
source. Recall that $n$ determines the fineness of the quantization
from $\mathcal{P}(\mathbb{X})$ to $\mathcal{P}_n(\mathbb{X})$.

\begin{theorem}[Discounted distortion] \label{theorem:near-optimal}
    For any $n\ge 1$ and
    $\beta\in (0,1)$, the sequence
    $\{\mathsf{Q}_t\}_{t\ge0}$ converges almost surely to a limit
    $\mathsf{Q}^*$. For any $\pi\in \mathcal{P}(\mathbb{X})$, let
    $\hat{\pi}$ denote the nearest neighbor of $\pi$ in
    $\mathcal{P}_n(\mathbb{X})$ and define the encoding policy
    $\gamma_{\beta,n}(\pi)$ by setting 
    \begin{align}
        \label{eq:q}
        \gamma_{\beta,n}(\pi) = \argmin_{Q \in
            \mathcal{Q}}\mathsf{Q}^*(\hat{\pi},Q).
    \end{align}
    Then, for any $\epsilon > 0$ and $\beta \in (0,1)$, there
    exists $N\ge 1$ such that
    \begin{align*}
        J_\beta(\zeta, \gamma_{\beta, n}) \le J^*_\beta(\zeta) + \epsilon
    \end{align*}
    for all $n \ge N$.
\end{theorem}

\emph{Remarks:}

\begin{itemize}
    \item[(a)] We note that $\mathsf{Q}^*(\pi,Q)$ is the expected value
        of the discounted cost obtained by a policy that takes ``action'' $Q$
        (i.e., uses the quantizer $Q$) at an initial state $\pi$, and then follows the
        optimal policy for the rest of the time. Due to the step
        where the minimum of  $\mathsf{Q}^*(\hat{\pi},Q)$ is considered
        instead of that of  $\mathsf{Q}^*(\pi,Q)$, the encoding policy $
            \gamma_{\beta,n}(\pi)$ is a piecewise constant function of the actual
        belief $\pi\in \mathcal{P}(\mathbb{X})$.
        
    \item[(b)] In the theorem we have used the limiting Q-value
        $\mathsf{Q}^*$ to derive the desired policy $\gamma_{\beta,n}$. Similar (probabilistic and in-expectation) bounds
        can be given if we  use  $\mathsf{Q}_t$ for finite (large)  $t$, at
        the expense of a more involved analysis. See \cite{SzepesvariAsymptotic, Qu} for details on finite-time analysis for Q-learning.
\end{itemize}

In the preceding theorem the discounted cost (distortion) is
considered, a quantity which has limited significance in source
coding.
The next theorem, which is the main result of this paper, shows that
the policy obtained in Theorem~\ref{theorem:near-optimal}, for $\beta$
close enough to $1$ and $n$ large enough is also a
near-optimal policy for the average cost (distortion) problem. As in
Theorem~\ref{theorem:near-optimal}, 
we assume that  $\{X_t\}_{t \ge 0}$ starts from the unique invariant
distribution $\zeta$.

\begin{theorem}[Average distortion] \label{theorem:discount-to-average2}
    Let $\epsilon > 0$. Then
    for   all sufficiently large $\beta \in (0,1)$, there exists
    $N(\beta) \ge 1$ such
    that
    \[
        J(\zeta,\gamma_{\beta,n}) \le J^*(\zeta) + \epsilon
    \]
    for all $n\ge N(\beta)$, where  $\gamma_{\beta,n}$ is the policy obtained in \eqref{eq:q} of
    Theorem~\ref{theorem:near-optimal}. 
\end{theorem}

\section{Quantized Q-learning}\label{section:Q-learning}
In this section,  we first introduce an MDP
formulation of the zero-delay coding problem, which has been studied
in depth in \cite{YukLinZeroDelay} and \cite{wood2016optimal}, and
describe our motivation for a reinforcement learning approach to solve
this MDP. Then we review recent results on a quantized Q-learning
algorithm in the context of a \emph{general} MDP, including the
necessary assumptions. Sections~\ref{section:unique-ergodicity} and
\ref{section:proofs} are then dedicated to proving that these
 assumptions hold for the zero-delay coding MDP, allowing us
to apply the quantized Q-learning algorithm. The resulting algorithm
(quantized Q-learning applied to the zero-delay coding MDP) is then
exactly our Algorithm 1.

\subsection{Zero-Delay Coding as a Markov Decision Process (MDP)}\label{section:optimal quantizers}
It has been shown in \cite{YukLinZeroDelay, wood2016optimal} that solving the zero-delay coding problem is equivalent to solving the MDP with state space $\mathcal{P}(\mathbb{X})$, action space $\mathcal{Q}$, transition kernel $P(d\pi' | \pi, Q)$ (which is induced by the update equation \eqref{eq:pi_update}), and cost function $c(\pi,Q)$ given in \eqref{eq:cost}. The following is a key property of this MDP.

\begin{definition}
    A transition kernel $P(dz' | z, u)$ is called \emph{weakly continuous} if
    \begin{equation*}
        \int f(z')P(dz' | z,u)
    \end{equation*}
    is continuous in $(z, u)$ for all continuous and bounded $f$. If an MDP has a weakly continuous transition kernel, we call the MDP \emph{weak Feller}.
\end{definition}

\begin{lemma}\label{lemma:weak}~\cite[Lemma
        11]{YukLinZeroDelay}. 
    For any $\gamma\in \Gamma_{\text{WS}}$, the  transition kernel $ P(d\pi_{t+1}|\pi_t,Q_t) $ is weakly continuous.
\end{lemma}

The MDP formulation of the zero-delay coding problem has many
analytical advantages. For example, it allows the use of dynamic
programming and value iteration methods to prove existence
results. However, this representation entails several limitations. In
particular, even though our original source $\{X_t\}_{t \ge 0}$
takes finitely many values, admits a unique invariant measure, and has
explicit transition probabilities given by $P(x' | x)$, the
MDP representation has $\{\pi_t\}_{t \ge 0}$ as its state process,
which takes values in the uncountable set $\mathcal{P}(\mathbb{X})$,
and it has an analytically complicated transition probability
$P(d\pi' | \pi, Q)$ induced by the update
equation~\eqref{eq:pi_update}. This makes actual implementation of
dynamic programming and value iteration for the computation of optimal
policies difficult.

In particular, a traditional approach to obtain an optimal policy for
the discounted cost problem is to use the value iteration algorithm
given by
\begin{align*}
    J_t(\pi) = \min_{Q \in \mathcal{Q}} \Bigl[ c(\pi, Q) + \beta \int_{\mathcal{P}(\mathbb{X})}J_{t-1}( \pi')P(d\pi' | \pi, Q) \Bigr],
\end{align*}
for $t \in\{ 1, 2, \ldots\}$, with $J_0(\pi) = 0$. It can be shown
that our MDP satisfies the conditions for the convergence 
$J_t(\pi_0) \to J^*_\beta(\pi_0)$ as $t \to \infty$ to hold, and the
actions obtaining the above minimum converge to an optimal
policy~\cite{wood2016optimal}.

It turns out however that actually computing this value function is
difficult; the values clearly cannot be computed directly for each
state as the state space is uncountable. An approach would be to
quantize the MDP via an approximate model whose solution is near-optimal for the original model (e.g.\ via~\cite[Theorem
    4.27]{SaLiYuSpringer}). For zero-delay quantization, such an
approximate model would require numerical simulations for the
computation of transition probabilities, as one would need to place a
probability measure on sets of probability measures. Thus, computing the above values is very difficult except in trivial
cases. This motivates the use of a reinforcement learning algorithm in
which the transition probabilities are not computed or estimated
explicitly.

\subsection{Quantized Q-learning}   \label{section:Q-learningC}
A common reinforcement learning algorithm for finding optimal policies
is Q-learning, in which empirical value functions are recursively
updated based on observed realizations of the state, action, and
cost. Such an algorithm is guaranteed to converge to an optimal policy
for the discounted cost problem, but only in situations where the
state and action spaces are finite (among other mild assumptions,
see~\cite{Watkins}), and thus it is not applicable in our case.

A solution is ``quantized'' Q-learning, where we approximate the
original MDP using an MDP with a finite state space, and run
Q-learning on this model. Recent work \cite{KSYContQLearning}
and~\cite{kara2021convergence} give conditions under which the
resulting policy is near-optimal for the original MDP.  We note that
such a quantization strategy is not limited to a Q-learning
approach. For example, \cite{SaYuLi15c} considers a quantized value iteration
approach, but while  this solves the issue of the uncountable state space,
one must still contend with the difficult state dynamics given by
$P(\, \cdot\,  | \pi, Q)$. Furthermore, in such a quantization procedure,
one must compute probability measures over the transition kernel
itself, which makes the problem even more challenging. A Q-learning
approach avoids all of this complexity by learning the values
empirically.

First, we state the assumptions that allow  for the
application of  quantized
Q-learning to a general MDP. In the proof of our main result, we will later show that these
assumptions hold for the zero-delay coding MDP. Consider an MDP with
$\mathsf{Z}$-valued state $\{Z_t\}_{t \ge 0}$,
$\mathsf{U}$-valued control $\{U_t\}_{t \ge 0}$, transition kernel
$P(dz'  | z, u)$ and cost function
$c : \mathsf{Z} \times \mathsf{U} \to [0, \infty)$

\begin{assumption}\label{assumption:MDP} The MDP has the following properties:
    \begin{enumerate}
        \item[(i)] The transition kernel $P(dz' | z,u)$ is weakly continuous.
        \item[(ii)] The cost function $c$ is continuous and bounded.
        \item[(iii)] The action space $\mathsf{U}$ is finite.
        \item[(iv)] The state space $\mathsf{Z}$ is a compact subset
            of a Euclidean space.
    \end{enumerate}
\end{assumption}

Let $\{B_i\}_{i=1}^N$ be a partition of $\mathsf{Z}$ into compact
subsets and let
$\mathsf{Y} \coloneqq \{y_1,\ldots,y_N\}$, where $y_i \in B_i$. We
define a \emph{quantizer} on $\mathsf{Z}$ as a mapping
$f : \mathsf{Z} \to \mathsf{Y}$, such that
\[ f(z) = y_i \quad \text{if} \; z \in B_i. \]

Note that when we first introduced quantizers in
Section~\ref{section:introduction}, we were considering quantizers of
the \emph{information source} $\{X_t\}_{t \ge 0}$. Although the idea
is the same here, we emphasize that we are now considering
quantization of the state space of an MDP.

We also define the maximum radius of the $B_i$:
\begin{equation}
\label{dinf-def}
d_\infty(\mathsf{Y},\mathsf{Z})  \coloneqq \max_{i=1,\ldots,N} \max_{z
  \in B_i}\|z-y_i\|.
\end{equation} 

We now introduce the quantized Q-learning algorithm. Here, we let
$y \in \mathsf{Y}$ and $Y_t \coloneqq f(Z_t)$. Also, we apply a
(possibly randomized) policy $\eta \coloneqq \{\eta_t\}_{t \ge 0}$, where
$\eta_t : \mathsf{Y}^{t+1} \times \mathsf{U}^t \to
    \mathsf{U}$. We refer to this policy as the \emph{exploration policy}. Finally, we define the $\mathsf{Q}$-factors
$\{\mathsf{Q}_t\}_{t \ge 0}$ and the learning rate
$\{\alpha_t\}_{t \ge 0}$, where both $\mathsf{Q}_t$ and
$\alpha_t$ are maps from $\mathsf{Y} \times \mathsf{U}$ to
$[0, \infty)$. Note that since $\mathsf{Y}$ and $\mathsf{U}$ are
finite, $\mathsf{Q}_t$ and $\alpha_t$ are tabular.

\medskip 

\noindent \underline{\textbf{Algorithm 2: Quantized Q-learning (General MDPs)}}\cite{KSYContQLearning}\label{algorithm:Q-learning}

\begin{algorithmic}[1]
    \State Initialize $Z_0$
    \For{$t\ge0$}
    \State $U_t \sim \eta_t(Y_{[0,t]}, U_{[0,t-1]})$
    \State Sample $Z_{t+1}$ according to $ P(dz'  | z, u)$
    \If{$(Y_t,U_t) = (y,u)$}
    \State $ \mathsf{Q}_{t+1}(y,u) = (1- \alpha_t(Y_t,U_t))\mathsf{Q}_t(Y_t,U_t) + \alpha_t(Y_t,U_t)[c(Z_t,U_t) + \beta \; \underset{v\in\mathsf{U}}{\text{min}} \; \mathsf{Q}_t(Y_{t+1},v)]$
    \Else
    \State $\mathsf{Q}_{t+1}(y,u) = \mathsf{Q}_t(y,u)$
    \EndIf
    \EndFor
\end{algorithmic}

As shown in \cite{KSYContQLearning} and stated in
Theorem~\ref{theorem:convergence} below, Algorithm~2 converges to the
optimum if  the following set of
assumptions, together with Assumption~\ref{assumption:MDP}, are
satisfied. 

\begin{assumption}\label{assumption:Q-learning} In Algorithm~2,  we have
    \begin{enumerate}
        \item[(i)] $\alpha_t(y,u) =
                \frac{1}{1 + \sum_{k=0}^t \mathbf{1}({(Y_k,U_k) = (y,u)})}$.
        \item[(ii)] Under $\eta$, each $(y,u) \in \mathsf{Y} \times \mathsf{U}$ is hit infinitely often almost surely.
        \item[(iii)] Under $\eta$, the state process $\{Z_t\}_{t\ge0}$ admits a unique invariant measure $\phi$.
    \end{enumerate}
\end{assumption}
\begin{theorem}\label{theorem:convergence}\cite[Corollary 11]{KSYContQLearning}
    Under Assumptions~\ref{assumption:MDP}
    and~\ref{assumption:Q-learning}, $\mathsf{Q}_{t+1}$ in Algorithm~2 converges to a limit $\mathsf{Q}^*$. Furthermore,
    consider the (deterministic and stationary) policy obtained through
    \begin{align*}
        \eta^*(z) = \argmin_{v \in \mathsf{U}}\mathsf{Q}^*(f(z), v).
    \end{align*}
    Then,
    \[
        \big| J_\beta(z, \eta^*)-
        J^*_\beta(z)\big| \to 0   \text{ \ as \  $d_\infty(\mathsf{Y},\mathsf{Z}) \to 0$}.
    \]
    for all $z \in \mathsf{Z}$. That is, the policy obtained
    by taking the minimizing actions of $\mathsf{Q}^*$ and then
    making it constant over the quantization bins is near-optimal for
    fine enough quantization.
\end{theorem}

Returning to our application, we want to apply this algorithm to the
MDP defined in Section~\ref{section:optimal quantizers}. Using the
notation of  this section, we have
$\mathsf{Z} = \mathcal{P}(\mathbb{X})$ and
$\mathsf{U} = \mathcal{Q}$, with the cost function~\eqref{eq:cost}
and transition kernel $P(d\pi' | \pi, Q)$. We let $\mathsf{Y} = \mathcal{P}_n(\mathbb{X})$ (recall \eqref{eq:Pn}) and show that
Assumptions~\ref{assumption:MDP} and~\ref{assumption:Q-learning} hold
for this setup. The most challenging  of these will be verifying that
Assumption~\ref{assumption:Q-learning} (iii) holds for our MDP; the next section is
dedicated to proving this. 

\section{Unique Ergodicity Under a Uniform Exploration Policy}\label{section:unique-ergodicity}
To show the desired result, we will need the following supporting
results from the literature of partially observed Markov processes
(POMPs). In this section, we assume that the quantizers
$\{Q_t\}_{t \ge 0}$ are chosen uniformly from $\mathcal{Q}$, as in
Algorithm 1, and call this the \emph{uniform policy}. Unless
explicitly stated otherwise, we assume that this uniform policy is
used, and so we omit $\gamma$ from the notation for probabilities and
expectations.

\subsection{Predictor and Filter Merging}
Recall the definition of the \emph{filter} in \eqref{eq:filter}:
\[ \overline{\pi}_t(A) \coloneqq P_{\pi_0}(X_t \in A | q_{[0,t]}).\]
The filter admits a recursion equation similar to~\eqref{eq:pi_update}
(see e.g.~\cite{Chigansky, Handel}). Note that
these recursions are dependent on the initialization of $ \pi_0 $,
also called the \emph{prior}. We denote the predictor (respectively,
filter) process resulting from the prior $ \pi_0 = \nu $ as
$ \{\pi_t^\nu \}_{t\ge0} $ (respectively,
$ \{\overline{\pi}_t^\nu \}_{t\ge0} $). A common problem in filtering theory is that of \emph{filter stability} (see e.g.\cite{Chigansky, Handel}), which essentially asks when the process $\{\overline{\pi}^\nu_t\}_{t \ge 0}$ is insensitive to the initialization $\nu$. This will be a crucial tool for proving unique ergodicity. We first provide some necessary definitions.

\begin{definition}
    Let $\nu, \mu \in \mathcal{P}(\mathbb{X})$. The total variation
    distance     is 
    \[ \|\nu -\mu \|_{TV} \coloneqq \sup_{\|f\|_\infty \le 1} \Bigl|\int f\,
        d\nu  - \int f\, d\mu \Bigr|, \]
    where the supremum is taken over all measurable $f:\mathbb{X}\to [-1,1]$.
\end{definition}
\begin{definition}\label{definition:TV-stable}
    We say that the predictor (respectively, filter) process is \emph{stable in total variation almost surely} if for any $ \mu, \nu, \kappa \in \mathcal{P}(\mathbb{X})$, we have that $P_\kappa$ almost surely,
    \[ \lim_{t \to \infty}\|\pi_t^\mu - \pi_t^\nu\|_{TV} = 0.\]
\end{definition}
We will use the following lemmas to deduce predictor stability.

\begin{lemma}\label{lemma:filterThenPred}
    If the filter process is stable in total variation almost surely, then the predictor process is stable in total variation almost surely.
\end{lemma}
\begin{IEEEproof}
    Consider the source transition kernel $P(x'|x)$. We have that $\pi^\mu_{t+1}(x') = \sum_x P(x'|x)\overline{\pi}^\mu_t(x)$. By a classic result of Dobrushin \cite{dobrushin1956central}, this implies that $\| 
\pi^\mu_{t+1} - \pi^\nu_{t+1} \|_{TV} \le \| \overline{\pi}^\mu_t - \overline{\pi}^\nu_t \|_{TV}$. The result follows.
\end{IEEEproof}

\begin{lemma}\label{lemma:positive-density}\cite[Corollary 5.5]{Handel}
    Let $\{A_t\}_{t\ge0}$ be a discrete-time Markov chain and $\{B_t\}_{t\ge0}$ be a stochastic process such that the $B_t$ are conditionally independent given $\{A_t\}_{t\ge0}$. Also assume $P(B_t | A_{[0,\infty)}) = P(B_t | A_t)$, and that $P(B_t | A_t)$ has the form
    \[ P(B_t \in B | A_t) = \int_B g(A_t, b)\psi(db), \]
    where $g(a,b)$ is a probability density with respect to the $\sigma$-finite measure $\psi$. If $g$ is strictly positive, and $\{A_t\}_{t \ge 0}$ is aperiodic and Harris recurrent (that is, it visits every state infinitely often with probability one \cite[Definition 3.1.3]{yuksel2020control}), then the filter $\overline{\pi}_t(A) \coloneqq P(A_t \in A | B_{[0,t]})$ is stable in total variation almost surely.
\end{lemma}

\begin{lemma}\label{lemma:filter-stable}
    Under the  uniform policy, the filter process $\{\overline{\pi}_t\}_{t\ge0}$ is stable in total variation almost surely.
\end{lemma}

\begin{IEEEproof}
    We apply Lemma \ref{lemma:positive-density} to $\{X_t\}_{t \ge 0}$ and $\{q_t\}_{t \ge 0}$. Note that under the uniform policy, $P(q_t | X_{[0,\infty)}) = P(q_t | X_t)$, and we have
    \begin{align*}
   \MoveEqLeft P(q_t = q| X_t = x) & \\
        = & \sum_{Q} P(q_t = q| X_t = x, Q_t = Q)P(Q_t = Q | X_t = x)                                                                          \\
                     = & \sum_{Q} \mathbf{1}(Q(x) = q)P(Q_t = Q | X_t = x) \\
                     = & \smashoperator[lr]{\sum_{\{Q : Q(x) = q\}}}P(Q_t = Q | X_t = x).
    \end{align*}
    where the second equality follows from the fact that
    $q = Q(x)$ is deterministic. Now, under the uniform policy, the quantizer $Q_t$ is chosen independently
    and randomly, so $P(Q_t = Q | X_t = x) = P(Q_t = Q)$. Thus we have
    \[P(q_t = q| X_t = x) = \smashoperator[lr]{\sum_{\{Q : Q(x) = q\}}}P(Q_t = Q).\]
    
    Since we are considering the set of all possible quantizers, for
    any $(x,q) \in \mathbb{X} \times \mathcal{M}$, the set
    $\{Q : Q(x) = q\}$ is nonempty, and under the uniform policy, $P(Q_t = Q) > 0$ for all $Q \in \mathcal{Q}$. Thus
    $P(q_t = q | X_t = x) > 0$ for all $(x,q)$. This implies the function $g$ in Lemma \ref{lemma:positive-density} is positive. Finally, note that $\{X_t\}_{t \ge 0}$ evolves independently of the encoding policy; it is always irreducible and aperiodic (thus, since $\mathbb{X}$ is finite, it is Harris recurrent and aperiodic). The result follows from Lemma \ref{lemma:positive-density}.
\end{IEEEproof}
Lemmas \ref{lemma:filterThenPred} and \ref{lemma:filter-stable} immediately imply the following:
\begin{corollary}\label{corollary:predictor-stable}
    Under the uniform policy, the predictor process
    $\{\pi_t\}_{t\ge0}$ is stable in total variation almost surely.
\end{corollary}

\begin{theorem}\label{theorem:invariant}
    Under the uniform policy,
    $ \{\pi_t\}_{t\ge0} $ admits a unique invariant measure.
\end{theorem}

\begin{IEEEproof} The proof slightly generalizes an argument presented
    in~\cite[Corollary 3]{Stettner}.  Throughout, we use the notation
    $ \nu(f) \coloneqq \int fd\nu $. Note that, under any
    Walrand-Varaiya type policy (and in particular, under the uniform policy), the processes $\{\pi_t\}_{t\ge0}$ and
    $ \{(X_t,\pi_t)\}_{t\ge0} $ are Markov. Let $T$ and $S$ be the
    transition operators associated with $\{\pi_t\}_{t\ge0}$ and
    $ \{(X_t,\pi_t)\}_{t\ge0} $, respectively. Recall that \( \zeta \)
    is the unique invariant measure of our source \( \{X_t\}_{t \ge 0}
    \).

    First note that since since the exploration policy is uniform, the
 induced transition kernel $P(d\pi' | \pi)$ itself it weakly continuous (i.e., $\int f(\pi')P(d\pi'|\pi)$ is continuous in $\pi$). Since every Markov process with a weakly continuous kernel
    on a compact state space admits an invariant measure (see, e.g.,
    \cite{Hernandez-Lerma2003}), $\{\pi_t\}_{t \ge 0}$ has an invariant measure.  Thus, we are left with proving
    uniqueness. Assume that
    $ m_1,m_2 \in \mathcal{P}(\mathbb{X} \times \mathcal{P}(\mathbb{X}))
    $ are two invariant measures for $ \{(X_t,\pi_t)\}_{t\ge0} $. Then
    their projections on $ \mathbb{X} $ are invariant for
    $ \{X_t\}_{t\ge0} $. Then, by unique invariance of $ \zeta $ we
    have
    \[ m_i(dx,d\nu) = P_{m_i}(d\nu | x)\zeta(dx).  \]
    
    We now  show that $ m_1(F) = m_2(F) $ for each $ F $ on a set
    of measure-determining functions, namely those such that
    $ F(x,\nu) = \phi(x)H(\nu(\phi_1),\ldots,\nu(\phi_l)) $, where
    $ \phi \in C(\mathbb{X}), \phi_1,\ldots,\phi_l \in
        C(\mathbb{X})$, $H$ is bounded and Lipschitz continuous with
    constant $ L_H $, and $ l \in \mathbb{Z}_+ $~\cite{Stettner}.
    
    By invariance we have, for $ i=1,2 $, that
    \[ m_i(F) = \smashoperator[l]{\int\limits_{\mathbb{X} \times \mathcal{P}(\mathbb{X})}} \frac{1}{T}\sum_{t=0}^{T-1}S^t F(x,\nu)P_{m_i}(d\nu | x)\zeta(dx). \]
    
    Thus,
    \begin{align*}
        &|m_1(F) - m_2(F)|                                                                                                                           \\
         & \le \int\limits_{\mathbb{X} \times \mathcal{P}(\mathbb{X}) \times \mathcal{P}(\mathbb{X})} \frac{1}{T}\sum_{t=0}^{T-1} |S^t F(x,\nu_1) - S^t F(x,\nu_2)| \\
         & \hspace{4cm} \cdot P_{m_1}(d\nu_1|x)P_{m_2}(d\nu_2|x)\zeta(dx) \\
         & \le L_H \|\phi\| \quad\quad
        \smashoperator[l]{\int\limits_{\mathbb{X} \times
                \mathcal{P}(\mathbb{X}) \times \mathcal{P}(\mathbb{X})}}
        \frac{1}{T}\sum_{t=0}^{T-1}\mathbf{E}[\sum_{i=1}^{l}|\pi_t^{\nu_1}(\phi_i)
        - \pi_t^{\nu_2}(\phi_i)|] \\
        & \hspace{4cm} \cdot P_{m_1}(d\nu_1|x)P_{m_2}(d\nu_2|x)\zeta(dx).
    \end{align*}
    Since the predictor process is stable in total variation almost surely,
    and by the dominated convergence theorem, the last integral  converges
    to zero as $ n \to \infty $. Then, the joint process
    $\{(X_t, \pi_t)\}_{t\ge0}$ admits at most one invariant
    measure. Next we show that $\{\pi_t\}_{t\ge0}$ admits at most
    one invariant measure.
    
    Assume that $ v_1,v_2 \in \mathcal{P}(\mathcal{P}(\mathbb{X})) $
    are two different invariant measures for $ \{\pi_t\}_{t\ge0}
    $. Then there exists a continuous and bounded $f : \mathcal{P}(\mathbb{X}) \to \mathbb{R}$ such that $v_1(f) \neq v_2(f)$. Now for $j = 1,2$,
    let $\{(x^j_t, \pi^j_t)\}_{t\ge0}$ be the process with initial law
    $\pi(dx)v_j(d\pi)$. Since $\mathbb{X}$ is finite, $P(X_t^j
        \in\,  \cdot\, , \pi_t^j \in \,  \cdot\, )$ is tight.
    
    Now, since $\mathbb{X}$ is finite,
    we also have that $\{(X_t, \pi_t)\}_{t\ge0}$ has a weakly
    continuous transition kernel. Thus the time average
    \[ \frac{1}{T}\sum_{t=0}^{T-1}P(x^j_t \in\,  \cdot\, , \pi^j_t \in
        \, \cdot\, ) \]
    converges weakly to an invariant measure $\eta_j$ for $\{(X_t, \pi_t)\}_{t\ge0}$~\cite[Theorem 3.3.1]{yuksel2020control}.
    
    Then for $F(x,\pi) = f(x)$, we have $\eta_1(F) = v_1(f) \neq v_2(f) = \eta_2(f)$. But then $\eta_1$ and $\eta_2$ are two different invariant measures for $\{(X_t, \pi_t)\}_{t\ge0}$, which is a contradiction. Thus $\{\pi_t\}_{t\ge0}$ admits at most one invariant measure.
    
\end{IEEEproof}

\subsection{Properties of the Unique Invariant Measure and Implications for Learning}\label{section:support}
Here we identify some properties of the unique invariant measure guaranteed by Theorem~\ref{theorem:invariant}, which will play a role in the use of Algorithm 2. Throughout, for any measure $\pi \in \mathcal{P}(\mathbb{X})$, we denote its nearest neighbor in $\mathcal{P}_n(\mathbb{X})$ by $\hat{\pi}$ (recall that this nearest neighbor map is performed using Algorithm 3).

Let $\eta$ be the uniform policy, and let $\phi$ be
the unique invariant measure for $\{\pi_t\}_{t \ge 0}$ under
$\eta$, as in Theorem 4. Not every element of $\mathcal{P}_n(\mathbb{X})$ (recall~\eqref{eq:Pn}) is hit infinitely often under
$\eta$. As a trivial example, take the case where
$\{X_t\}_{t \ge 0}$ is independent and identically distributed
(i.i.d.) with distribution given by $\pi$. Then we have
$\pi_t(x) = \pi(x)$ for all $t \ge 1$, so that only $\hat{\pi} \in \mathcal{P}_n(\mathbb{X})$ is
visited infinitely often.

To address this issue, consider the set $\mathcal{B}^\phi \coloneqq \{B \in \{B_i\}_{i=1}^N : \phi(B_i) > 0\}$ where $\{B_i\}_{i=1}^N$ is the set of bins of $\mathcal{P}(\mathbb{X})$ under the nearest neighbor map. Then consider the set of all $\hat{\pi} \in \mathcal{P}_n^{\phi}(\mathbb{X})$ whose corresponding bin has positive measure under $\phi$, given by
\begin{equation}
    \mathcal{P}_n^{\phi}(\mathbb{X}) \coloneqq \left\{ \hat{\pi} \in \mathcal{P}_n(\mathbb{X}) : \hat{\pi} \in B \text{ for some } B \in \mathcal{B}^\phi \right\}\label{eq:PnPhi}.
\end{equation}
\begin{lemma}\label{lemma:infinite_hits}
    Under the uniform policy $\eta$, for every $\hat{\pi} \in \mathcal{P}_n^\phi(\mathbb{X})$, we have $\hat{\pi}_t = \hat{\pi}$ infinitely often almost surely.
\end{lemma}

\begin{IEEEproof}
    By the pathwise ergodic theorem (e.g., \cite[Theorem 5.4.1]{Hernandez-Lerma2003}) there exists some $\mu \in \mathcal{P}(\mathbb{X})$ such that for all measurable and bounded $g : \mathcal{P}(\mathbb{X}) \to \mathbb{R}$,
    \[ \frac{1}{N} \sum_{t=0}^{N-1} g(\pi_t^\mu) \to \int g(\pi)\phi(d\pi) \]
    $P_\mu$ almost surely as $N \to \infty$. But by Corollary~\ref{corollary:predictor-stable}, this implies that
    \begin{equation}
    \frac{1}{N} \sum_{t=0}^{N-1} g(\pi_t^\nu) \to \int g(\pi)\phi(d\pi) \label{eq:limit_pred_stability}
    \end{equation}
    $P_\kappa$ almost surely for all $\nu, \kappa$. Now let $g$ be the indicator function of the quantization bin of the nearest neighbor map $f$ corresponding to some $\hat{\pi} \in \mathcal{P}_n^\phi(\mathbb{X})$; that is, $g(\pi) = 1$ if $f(\pi) = \hat{\pi}$ and $g(\pi) = 0$ otherwise. Then~\eqref{eq:limit_pred_stability} implies that any bin which has positive measure under $\phi$ must be hit infinitely often almost surely by $\pi^\nu_t$. But this is exactly how we defined $\mathcal{P}_n^\phi(\mathbb{X})$; it is the set of all $\hat{\pi}$ whose bins have positive measure under $\phi$. Therefore every $\hat{\pi} \in \mathcal{P}_n^\phi(\mathbb{X})$ is hit infinitely often almost surely. Note that $\nu$ was arbitrary, so this holds regardless of the initialization $\pi_0 = \nu$.
\end{IEEEproof}

Now that we have identified a set $\mathcal{P}^\phi_n(\mathbb{X})$ that is hit infinitely often (and thus, one where Assumption~\ref{assumption:Q-learning} (ii) is valid), we claim that we can restrict ourselves to this set without any loss of optimality. This is formalized in the following lemma and corollary.
\begin{lemma}\label{lemma:reachable}
    Let $\pi_0 \sim \kappa$, where $\kappa \ll \phi$ ($\kappa$ is absolutely continuous with respect to $\phi$) and let $\gamma \in \Gamma_{WS}$. Then for all $\hat{\pi}$ that are reachable from $\hat{\pi}_0$ under $\gamma$ (that is, such that $P^\gamma(\hat{\pi}_t = \hat{\pi} | \pi_0 \sim \kappa) > 0$ for some $t$), we have $\hat{\pi} \in \mathcal{P}^\phi_n(\mathbb{X})$.
\end{lemma}
\begin{IEEEproof}
     By invariance of $\phi$ under the uniform policy, we have that
     for any $t \ge 0$,
\begin{multline*}   
\phi(B’) = \frac{1}{t} \sum_{k=0}^{t-1} \sum_{\overline{Q} \in \mathcal{Q}^t} \frac{1}{|\mathcal{Q}^t|} \mathbf{1}(Q_{[0,t-1]}= \overline{Q})  \\ \cdot \int \phi(d\pi) P(\pi_k \in B' | \pi_0 = \pi, Q_{[0,t-1]} = \overline{Q}),
 \end{multline*} 
where $B'$ is the bin of $\hat{\pi}'$, and the second sum is over all $\overline{\mathcal{Q}} \in \mathcal{Q}^t$, i.e. over every possible realization of $Q_{[0,t-1]}$. Now assume that $\kappa(A) = \frac{\phi(A)}{\phi(B)}$ for all $A \subset B$, where $B$ is some bin of the nearest neighbor map with $\phi(B) > 0$; that is, $\kappa$ is the restriction of $\phi$ on $B$. Then given any sequence of quantizers $Q_{[0,t-1]} = \overline{Q}$, we have
	\begin{align*}
	 \MoveEqLeft[6] P(\hat{\pi}_t = \hat{\pi}' | \pi_0 \sim \kappa, Q_{[0,t-1]} = \overline{Q}) \\
     = & P(\pi_t \in B' | \hat{\pi}_0 = \hat{\pi}, Q_{[0,t-1]} = \overline{Q})\\
     = & \int \kappa(d\pi) P(\pi_t \in B' | \pi_0 = \pi, Q_{[0,t-1]} = \overline{Q}).
	\end{align*} 
Since the above holds for any sequence of quantizers, for any $\gamma \in \Gamma_{WS}$ we have 
	\[P^\gamma(\hat{\pi}_t = \hat{\pi}' | \pi_0 \sim \kappa) \ll \phi(B').\]
Furthermore, the above holds for any $\kappa \ll \phi$ since the order of absolute continuity holds. Thus any reachable $\hat{\pi}$ must be such that its bin $B$ satisfies $\phi(B) > 0$, and thus $\hat{\pi} \in \mathcal{P}^\phi_n(\mathbb{X})$.
\end{IEEEproof}

\begin{corollary}\label{corollary:reachableFromSupport}
    Let $\pi_0 = \pi^*$, where $\pi^* \in \text{supp}(\phi)$ and $\text{supp}(\phi)$ denotes the support of $\phi$ in $\mathcal{P}(\mathbb{X})$, and let $\gamma \in \Gamma_{WS}$. Then for all $\pi$ that are reachable from $\pi_0$ under $\gamma$, we have either (i) $\pi \in B$ for some $B \in \mathcal{B}^\phi$, or (ii) $\pi$ is on the boundary of some $B \in \mathcal{B}^\phi$.
\end{corollary}
\begin{IEEEproof}
    Let $\{N_m\}_{m \ge 0} \subset \mathcal{P}(\mathbb{X})$ be a sequence of open balls centered at $\pi^*$ such that $\bigcap_{m=0}^\infty N_m = \pi^*$ and let $\{\kappa_m\}_{m \ge 0} \subset \mathcal{P}(\mathcal{P}(\mathbb{X}))$ be defined as $\kappa_m(A) = \frac{\phi(A)}{\phi(N_m)}$ for all $A \subset N_m$; that is, $\kappa_m$ is the restriction of $\phi$ to $N_m$. Note that by definition, $\phi(N_m) > 0$ and $\kappa_m \ll \phi$ for all $m$. We also have by weak continuity of $P(d\pi' | \pi, Q)$ that $P(d\pi_t | \pi_0 \sim \kappa_m, Q_{[0,t-1]} = \overline{Q})$ converges weakly to $P(d\pi_t | \pi_0 = \pi^*, Q_{[0,t-1]})$.
    
    Now let $B' \subset \mathcal{P}(\mathbb{X})$ be open. By the Portmanteau theorem (e.g., \cite[Theorem 11.1.1]{Dudley}), we have that $\liminf_{m \to \infty}P(\pi_t \in B' | \pi_0 \sim \kappa_m, Q_{[0,t-1]}) \ge P(\pi_t \in B' | \pi_0 = \pi^*, Q_{[0,t-1]})$. By the same argument as the previous lemma, this implies that $P^\gamma(\pi_t \in B' | \pi_0 = \pi^*) \ll \phi(B')$ for any $\gamma \in \Gamma_{WS}$. 
    
    Now assume that $\pi$ is reachable, and take some open ball $N_\epsilon(\pi)$ around $\pi$. The above argument implies that $\phi(N_\epsilon(\pi)) > 0$. Since this holds for arbitrary $\epsilon$, it must be that either (i) or (ii) holds.
    
\end{IEEEproof}
To summarize, the previous lemma and corollary tell us the following:
\begin{enumerate}
    \item If our initial measure $\pi_0$ is sampled according to a distribution which is absolutely continuous with respect to $\phi$, then with probability one the $\{\hat{\pi}_t\}_{t \ge 0}$ process will stay in $\mathcal{P}^\phi_n(\mathbb{X})$.
    \item If instead our initial measure $\pi_0$ is deterministically set to an element of the support of $\phi$, then with probability one the $\{\pi_t\}_{t \ge 0}$ process will stay in $\mathcal{P}^\phi_n(\mathbb{X})$ up to the tie breaking on decision boundaries of $\mathcal{P}_n(\mathbb{X})$.
\end{enumerate}

In practice, the decision boundaries of $\mathcal{P}_n(\mathbb{X})$ are known to us, and furthermore the elements of $\mathcal{P}^\phi_n(\mathbb{X})$ will also be clear after running Algorithm 1 for a sufficiently long time, as these will be the bins that are hit most often. Accordingly, one can easily modify the decision boundaries such that any reachable $\hat{\pi}$ is in $\mathcal{P}^\phi_n(\mathbb{X})$. We assume that such a modification has been done, and henceforth restrict our near-optimal policies to only $\mathcal{P}^\phi_n(\mathbb{X})$. Finally, we show that one such initialization for $\pi_0 \in \text{supp}(\phi)$ is given by $\zeta$, the invariant distribution of the source.
\begin{lemma}\label{lemma:support}
  We have  $\zeta \in \text{supp}(\phi)$, where  $\text{supp}(\phi)$
  denotes the support of $\phi$ in $\mathcal{P}(\mathbb{X})$.
\end{lemma}

\begin{IEEEproof}
    Consider some open neighborhood of radius $\delta$ containing $\zeta$, say $N_\delta(\zeta) \subset \mathcal{P}(\mathbb{X})$. Now consider a totally ``uninformative'' quantizer $Q \in \mathcal{Q}$; that is $Q(x) = i$ for all $x \in \mathbb{X}$ and some $i \in \mathcal{M}$. Via the update equation (4), if $Q_t = Q$, we have that $\pi_{t+1} = \pi_t P$, where we have used the matrix notation $P(i,j) \coloneqq P(X_{t+1} = j | X_t = i)$. By a classical result of Dobrushin~\cite{dobrushin1956central}, there exists some $T > 0$ such that for all $\pi \in \mathcal{P}(\mathbb{X})$, $\pi P^T \in N_\delta(\zeta)$. Thus, if $Q_t =$ for $t = 0,\ldots,T-1$, then we have $\pi_T \in N_\delta(\zeta)$.
    
    But under the uniform policy, we have some fixed positive
    probability of choosing $Q$, say $P(Q_t = Q) = p > 0$. In
    particular, for all $t \ge T$,
    $P(\pi_t \in N_\delta(\zeta) | \pi_0 = \pi) \ge P(Q_{[t-T,t-1]} =
    (Q, Q, \ldots, Q)) = p^T$. This implies that $\zeta$ is
    ``accessible'' (see~\cite[Definition 2.1]{Hairer}) and hence that
    $\zeta \in \text{supp}(\phi)$~\cite[Lemma 2.2]{Hairer}.
\end{IEEEproof}

\section{Proofs of Main Results} \label{section:proofs}
We are now ready to prove Theorems~\ref{theorem:near-optimal} and \ref{theorem:discount-to-average2}
with the aid of the auxiliary results developed in  Sections~\ref{section:Q-learning} and
\ref{section:unique-ergodicity}. In particular, we show that  Assumptions~\ref{assumption:MDP}
and~\ref{assumption:Q-learning} hold for our quantized MDP with
components 
\begin{gather}
    \mathsf{Z} = \mathcal{P}(\mathbb{X}), \; \mathsf{Y} =
    \mathcal{P}^\phi_n(\mathbb{X}), \; \mathsf{U} = \mathcal{Q}, \label{eq:quantized-MDP} \\ P = P(\,
    \cdot\,  | \pi, Q), \; c = c(\pi, Q) \nonumber,
\end{gather}
from which the proofs will follow. We also let
$d_\infty= d_\infty(\mathsf{Y},\mathsf{Z})$ for notational simplicity,
where  $d_\infty(\mathsf{Y},\mathsf{Z})$ was defined in
\eqref{dinf-def}.

Let $f$ map $\pi \in  \mathcal{P}(\mathbb{X})$ to its nearest neighbor
$\hat{\pi} \in  \mathcal{P}_n(\mathbb{X}) \subset
    \mathcal{P}(\mathbb{X})$. Note that we consider the smaller set $\mathcal{P}^\phi_n(\mathbb{X})$ rather than all of $\mathcal{P}_n(\mathbb{X})$, but this is without any loss when we start from $\pi_0 = \zeta$ by Corollary~\ref{corollary:reachableFromSupport} and Lemma~\ref{lemma:support}.

\begin{lemma}\label{lemma:assumptions}
    The (quantized) MDP defined by the components in~\eqref{eq:quantized-MDP} satisfies Assumptions~\ref{assumption:MDP} and~\ref{assumption:Q-learning}.
\end{lemma}
\begin{IEEEproof}
    We begin with Assumption~\ref{assumption:MDP}: (i) holds by
    Lemma~\ref{lemma:weak}. (ii) is clear from our definition of $c$
    in~\eqref{eq:cost}. Since $\mathbb{X}$ and $\mathcal{M}$ are
    finite, $\mathcal{Q}$ is finite, so (iii) holds. Finally, 
    $\mathbb{X}$ is finite, so $\mathcal{P}(\mathbb{X})$ is
    compact, so (iv) holds.
    
    We now show Assumption~\ref{assumption:Q-learning}: (i)
    holds in Algorithm 1 based on the update
    equation~\eqref{eq:alpha_update}. (ii) holds from Lemma~\ref{lemma:infinite_hits} and from the uniform selection of
    $Q_{t+1} \in \mathcal{Q}$. Finally, (iii) holds due to
    Theorem~\ref{theorem:invariant}.
\end{IEEEproof}

We can now prove our main results.

\begin{IEEEproof}[Proof of Theorem~\ref{theorem:near-optimal}]
  Algorithm~1 is simply the quantized Q-learning algorithm in
  Section~\ref{section:Q-learning} applied to the quantized zero-delay
  coding MDP in~\eqref{eq:quantized-MDP}, and using the uniform policy
  for $\eta$. Note that by definition of $\mathcal{P}_n(\mathbb{X})$,
  for any $\pi=(p_1,\ldots,p_m)\in  \mathcal{P}(\mathbb{X})$ and
  corresponding   $\hat{\pi} = (\hat{p}_1,\ldots,\hat{p}_m)\in
  \mathcal{P}_n(\mathbb{X})$, we have 
  $|p_i - \hat{p}_i| \le \frac{1}{n}$ for  $i=1,\ldots,n$, so the
  maximum radius of the quantization bins satisfies
    \[
        \max_{\pi \in \mathcal{P}(\mathbb{X})} \min_{\hat{\pi} \in
            \mathcal{P}_n(\mathbb{X})} \|\pi -\hat{\pi}\|_2 = O\Bigl(\frac{1}{n}\Bigr) \to 0, 
    \]
    so we have $d_\infty \to 0$ as $n \to \infty$. Then by Lemma~\ref{lemma:assumptions} we can apply
    Theorem~\ref{theorem:convergence} when $\pi_0 = \zeta$. Thus for any $\beta\in (0,1)$ and
    $\epsilon>0$ we can choose $N$ such that for all $n\ge N$ in
    Algorithm~1,
    \begin{equation}
        \label{eq:disc-bound}
        J_\beta(\zeta, \gamma_{\beta, n}) \le J^*_\beta(\zeta) + \epsilon
    \end{equation}
    as claimed.
\end{IEEEproof}

\begin{IEEEproof}[Proof of Theorem~\ref{theorem:discount-to-average2}]
    We will use Theorem~\ref{theorem:discounted-near-optimal} in
    Appendix \ref{appendix:A} that, loosely speaking, states that if a policy is near-optimal for a discount factor $\beta$ close enough to $1$, then this
    policy is also near-optimal for the average cost.  Let us verify
    that the conditions of the theorem are met when
    $\mathsf{Z} = \mathcal{P}(\mathbb{X})$,
    $\mathsf{U} = \mathcal{Q}$,
    $P(\, \cdot\,  | z, u) = P(\, \cdot\,  | \pi, Q)$, and $c(z,u) =
        c(\pi,Q)$. Indeed, (i)-(iii) are met by the definition of
    $c(\pi, Q)$ since  $\mathcal{P}(\mathbb{X})$ is the
    standard probability simplex in $\mathbb{R}^{|\mathbb{X}|}$  and
    $\mathcal{Q}$ is finite. Condition (iv) is true by Lemma~\ref{lemma:weak},
    and (v) is holds  by~\cite[Lemma 1]{wood2016optimal}, which states
    that for any $\pi, \pi' \in \mathcal{P}(\mathbb{X})$,
    \begin{equation*}
        \left| J^*_\beta(\pi) - J^*_\beta(\pi') \right| \le K
        \rho_1(\pi, \pi') \le K |\mathbb{X}|, 
    \end{equation*}
    where $K$ is some constant and $\rho_1$ is the $L_1$ Wasserstein
    distance on $\mathcal{P}(\mathbb{X})$~\cite{villani2008optimal}.
    
    Now let $\epsilon>0$ and let $N(\beta) \ge 1$ be such that 
    the policy $\gamma_{\beta,n}$ in
    Theorem~\ref{theorem:near-optimal} satisfies
    $ J_\beta(\zeta, \gamma_{\beta, n}) \le
        J^*_\beta(\zeta) + \frac{\epsilon}{2(1-\beta)}$ for all $n\ge N(\beta)$. Then we can apply
    Theorem~\ref{theorem:discounted-near-optimal}  with $\epsilon$
    replaced by $\epsilon/2$ and $\delta$ replaced by
    $\frac{\epsilon}{2(1-\beta)}$ to conclude that for all 
    $\beta\in (0,1)$  close enough to $1$,
    the average cost performance of  $\gamma_{\beta,n}$  satisfies 
    \[
        J(\zeta, \gamma_{\beta, n}) \le J^*(\zeta) + \epsilon
    \]
    for all $n\ge N(\beta)$, 
    which completes the proof. 
\end{IEEEproof}

\emph{Remark:} Note that due to the way the discounted cost scales with
$\beta$, the policy $\gamma_{\beta,n}$ in the proof does not have to
be $\epsilon$-optimal for the discounted cost,  but rather it can be
chosen to be only $\frac{\epsilon}{2(1-\beta)}$-optimal.

\section{Simulations}\label{section:results}
In the simulations we use mean-squared error (MSE)   \(
d(x,\hat{x}) = (x - \hat{x})^2 \)  as our distortion
measure and when using
Algorithm 1 we fix the discount factor at \(\beta = 0.9999\). We let $\pi_0 = \zeta$, so that Theorem~\ref{theorem:discount-to-average2} is applicable. When
testing algorithm performance, we take the average distortion over
\(10^6\) samples, which we denote by \(D\), and calculate the
signal-to-noise ratio (SNR) according to
\begin{equation*}
    \text{SNR} = 10\log_{10} \left( \frac{\text{var}(X)}{D} \right),
\end{equation*}
where \(\text{var}(X)\) is the variance of the stationary
Markov source \(\{X_t\}_{t \ge
0}\). We plot the SNR for varying quantizer rates, which is given by
\(R=\log_2(|\mathcal{M}|)\). Finally, in all of the simulations we
take \(\hat{\mathbb{X}} = \mathbb{X}\subset \mathbb{R}\).

\subsubsection*{On algorithm complexity and time to convergence} Note
the set \(\mathcal{P}_n(\mathbb{X})\), and thus the state space of our
quantized MDP, has size
\( \left| \mathcal{P}_n(\mathbb{X}) \right| = \)
\({n+|\mathbb{X}|-1}\choose{|\mathbb{X}|-1}\)~\cite{Reznik}. Furthermore,
our action space, if we include every possible quantizer from
\(\mathbb{X} \to \mathcal{M}\), has size
\(\mathcal{Q} = |\mathcal{M}|^{|\mathbb{X}|}\). Although these both
grow quickly in their respective parameters, we note that the actual
utilized state and action spaces tend to be much smaller. Indeed, $\mathcal{P}_n^\phi(\mathbb{X})$ tends to be much smaller than $\mathcal{P}_n(\mathbb{X})$; as an
extreme case, for an independent and identically distributed (i.i.d.)
source, $\mathcal{P}_n^\phi(\mathbb{X})$ has only one element. Also, many quantizers do not need
to be considered (for example, those with empty bins, and for certain
distortion measures those with non-convex bins). Thus, the actual
number of states and convergence time will generally be much less than
the theoretical upper bound. Note that for linear learning rates (as
in our algorithm), the required number of iterations for convergence
is polynomial in \(|\mathcal{P}_n(\mathbb{X})| \times |\mathcal{Q}|\)
(see~\cite{even2004learning} for related bounds on convergence time
for Q-learning).

\subsection{Finite-Alphabet Markov Sources}
For finite-state Markov sources, we compare Algorithm 1 against an
omniscient finite-state scalar quantizer (O-FSSQ)~\cite[Chapter
    14]{GershoVectorQuant} as this method seems to be the most competitive
among zero-delay code designs for Markov sources. In this algorithm,
one fixes a quantizer \(V\) from \(\mathbb{X}\) to
\(\{1, \ldots, K\}\), where \(\{1, \ldots, K\}\) is a finite set of
``states''. A sequence of training data is sorted into subsequences
based on the state of the preceding data point. A Lloyd-Max
quantizer~\cite{LloydIT82} is then designed for each of these
subsequences. However, in implementation the decoder does not know
exactly the state of the preceding data point, so it uses
\(V(\hat{X}_t)\), rather than \(V(X_t)\), as the state. That is, if
\(V(\hat{X}_t) = i \in \{1, \ldots, K\}\), then to quantize
\(X_{t+1}\) one uses  the Lloyd-Max quantizer corresponding to the
$i$th  subsequence. The O-FSSQ is known to perform very well for
many common Markov sources and on real-world sampled data.

The O-FSSQ employs a strategy  similar to Algorithm 1 in the sense
that it uses a finite set of ``states'' at time \(t-1\) to select a
quantizer at time \(t\). While simple to implement and relatively fast
to train, the O-FSSQ design is based on heuristic principles and has
no guarantee for optimality or near-optimality, unlike our 
Algorithm 1. Note that since \(\hat{\mathbb{X}}\) is finite, the
O-FSSQ is limited to using at most \(K=|\hat{\mathbb{X}|}\)
states. On the other hand, in Algorithm 1 we can let the number of states be
larger in order to obtain a better performance.

\begin{figure}[H]
    \centering
    \includegraphics[scale=0.59]{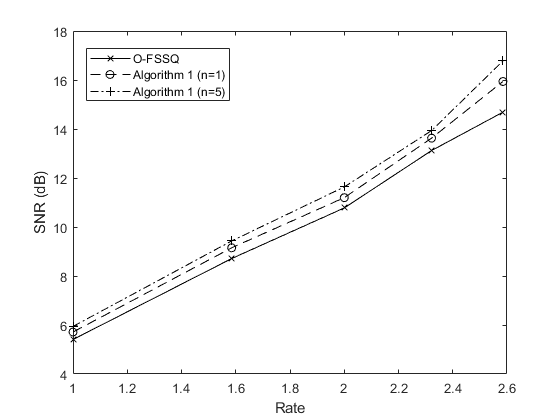}
    \caption{Comparison of Algorithm 1 with O-FSSQ for finite Markov
        source}
    \label{fig1}
\end{figure}

In the simulations we consider a stationary Markov source with common
source and reproduction alphabet
\(\mathbb{X} = \hat{\mathbb{X}} = \{1, \ldots, 8\}\). The transition
matrix and the unique invariant distribution are given below. The
transition matrix was randomly (uniformly)  chosen from the set of all
$|\mathbb{X}|\times |\mathbb{X}|$  transition
matrices (in which case the invariant distribution is unique with
probability one). In Fig.~\ref{fig1} we plot SNR values for
\(|\mathcal{M}| = 2, \ldots, 6\), so the rates range from $R=1$ to
$R= \log_2(6)$. We include an O-FSSQ designed with \(K=8\) states,
using \(10^6\) training samples. We include two comparisons with
Algorithm 1. The first is with \(n=1\), which yields \(8\) states. The
second is with \(n=5\), which gives \({12}\choose{7}\)\(=792\)
possible states, but at most \(30\) are actually utilized in the final
design.

\medskip 

\subsubsection*{Transition matrix of source in  Figure 1}
\footnotesize
\begin{equation*}
\setlength\arraycolsep{2pt}
    \begin{pmatrix}
        0.1331 & 0.0824 & 0.0311 & 0.2131 & 0.2623 & 0.0714 & 0.0417 & 0.1645  \\
        0.1207 & 0.1501 & 0.1268 & 0.1974 & 0.0952 & 0.0862 & 0.1870 & 0.0362  \\
        0.2320 & 0.0491 & 0.1770 & 0.1476 & 0.1530 & 0.1691 & 0.0215 & 0.05043 \\
        0.0162 & 0.1930 & 0.2511 & 0.1935 & 0.0688 & 0.1280 & 0.0893 & 0.0597  \\
        0.0420 & 0.1496 & 0.1130 & 0.0478 & 0.1073 & 0.2345 & 0.0692 & 0.2363  \\
        0.1382 & 0.1720 & 0.1378 & 0.1369 & 0.0396 & 0.1923 & 0.1383 & 0.0445  \\
        0.1710 & 0.2153 & 0.1579 & 0.0366 & 0.1530 & 0.1144 & 0.0439 & 0.1075  \\
        0.1292 & 0.0534 & 0.1309 & 0.0315 & 0.2837 & 0.2617 & 0.0103 & 0.0988
    \end{pmatrix}.
\end{equation*}
\normalsize
\subsubsection*{Invariant distribution}
\footnotesize
\begin{equation*}
\setlength\arraycolsep{2pt}
    \begin{pmatrix}
        0.1211 & 0.1326 & 0.1416 & 0.1328 & 0.1360 & 0.1580 & 0.0806 & 0.0973
    \end{pmatrix}.
\end{equation*}
\normalsize
\vspace{-1em}

Note that even when using the same number of states, Algorithm~1
outperforms the O-FSSQ. At rate $R=1$, the SNR gain over the O-FSSQ is
0.31 dB for $n=1$ and 0.53 dB for $n=5$.  At rate $R=\log_2(6)=
    2.58$, the SNR gain over the
O-FSSQ is 1.25 dB for \(n=1\) and 2.09 dB for \(n=5\). Moreover, Algorithm~1 can be used with a
larger number of states, giving performance gains. Of course, this
comes at a cost of overall codebook size, as a different
quantizer/codebook must be stored for each different state. 

\subsection{Continuous Markov Sources}
While the mathematical analysis presented in the paper does not cover
the continuous case, as future work we intend to develop a rigorous
treatment of the continuous source setup. The weak Feller and
ergodicity results follow as before, building on the
analysis in \cite{YukLinZeroDelay} that used the convex analytic method
to obtain structural results for continuous space Markov models;  however the
unbounded cost will require additional analysis. 
Nonetheless, in this section, we demonstrate that the algorithm is
also suitable for continuous sources as quantization of the
probability measures can be done in a variety of efficient ways,
facilitating reinforcement
learning.

Notably, \cite{kreitmeier2011optimal} and~\cite[Section
    V.C]{SYLTAC2017POMDP} propose a scheme in which a probability measure
is first approximated by one with finite support, then further
quantized using Algorithm 3. Under certain assumptions, this was shown
to be an efficient method for quantizing probability measures under a
Wasserstein metric, and consequently, the weak convergence topology.

In particular for our algorithm, one computes
\(\hat{\pi}_t \in \mathcal{P}_n(\mathbb{X})\) by first approximating
\(\pi_t \in \mathcal{P}(\mathbb{X})\) by a compactly-supported
measure, then approximating this by a finitely-supported one, and
finally applying Algorithm 3 to this measure. One similarly
approximates the space of quantizers \(\mathcal{Q}\) by  quantizers
on some finite set.

\begin{figure}[H]
    \centering
    \includegraphics[scale=0.59]{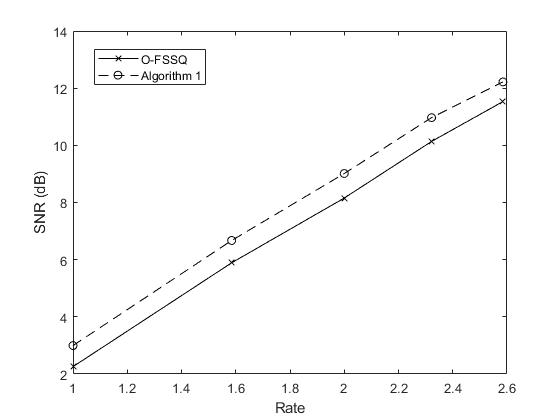}
    \caption{Comparison of Algorithm 1 with O-FSSQ for Gauss-Markov
        source}
    \label{fig2}
\end{figure}

In the simulations, we apply this strategy to a Gauss-Markov source
with correlation coefficient of 0.9, and we again consider
\(|\mathcal{M}| = 2, \ldots, 6\). For quantization of
\(\mathcal{P}(\mathbb{X})\) and \(\mathcal{Q}\), we approximate
\(\mathbb{X} = \mathbb{R}\) by the finite set
$\{-6+0.05i: i=0,1,\ldots,240\}$ and use \(n=5\) in Algorithm~1. This
results in no more than \(70\) states visited for any given rate, so
we compare it to an O-FSSQ using \(K=70\) states, trained on \(10^6\)
samples, where we use a Lloyd-Max quantizer for the state classifier
\(V\). Note that in this case the number of states \(K\) for the
O-FSSQ is not limited, so we only provide a comparison for the same
number of states. The results are shown in Fig.~\ref{fig2}. At rate
$R=1$, the SNR gain over the O-FSSQ is 0.74 dB and at rate $R=2.58$, the SNR
gain over the O-FSSQ is 0.69 dB.

\vspace{-1em}

Note that the question of near-optimality in the
continuous case is more intricate, since it depends not only on the
parameter \(n\) but also on the compact and finite approximations of
the support of \(\pi_t\). That is, even if \(n\) is large, Algorithm 1
may perform poorly if the finite approximation is not fine enough.

\subsection{Memoryless sources}
\begin{figure}[H]
    \centering
    \includegraphics[scale=0.59]{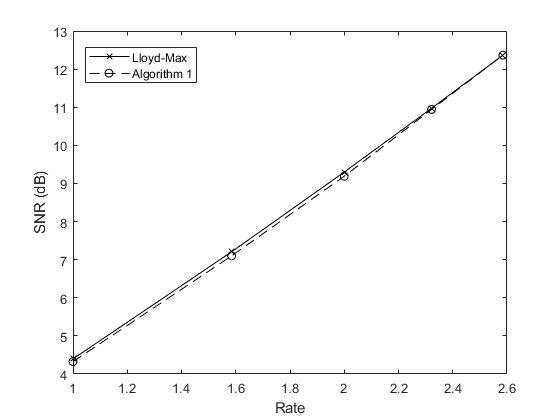}
    \caption{Comparison of Algorithm 1 with Lloyd-Max for
        i.i.d. Gaussian source}
    \label{fig3}
\end{figure}

Finally, to demonstrate the effectiveness of the (modified)
Algorithm~1, we compare it to the Lloyd-Max quantizer for continuous
i.i.d. sources. For such sources, a Lloyd-Max quantizer is only
guaranteed to be locally optimal \cite{LloydIT82}. On the other hand,
our Algorithm 1 converges to a globally optimal solution (as
$n\to \infty$) so it may outperform a Lloyd-Max quantizer also in the
i.i.d. case.  However, if the source has a log-concave density, then
all local optima coincide with the unique globally optimal
quantizer~\cite{kieffer1983uniqueness}, so a Lloyd-Max quantizer
designed using the source distribution yields the optimal solution
(which is the optimal zero-delay code for the i.i.d.\ source). To
verify near-optimality in this case, we compare Algorithm 1 (with the
same quantization parameters as the previous subsection) with a
Lloyd-Max quantizer designed using the source distribution, for a
standard Gaussian source \(X_t \sim N(0,1)\). The results are shown in
Fig.~\ref{fig3}. Here the Lloyd-Max design marginally outperforms
Algorithm~1 because of the approximation steps used in the latter in
the continuous case. The maximum SNR difference across all rates is
0.10~dB, occurring at rate $R=1.58$. At rate $R=2.58$, the difference
is only 0.002~dB.

\vspace{-1em}

\section{Conclusion}
\label{sec-concl}
We presented a reinforcement learning based algorithm for the
design of zero-delay codes for Markov sources. For finite alphabet
sources we proved that our Q-learning based
algorithm produces zero-delay codes that are optimal as the quantization of the underlying probability space becomes arbitrarily fine, provided the source starts from its invariant distribution. As far as we know, this
is the first provably optimal design method for zero-delay coding.
The performance of the algorithm was also demonstrated via
simulations for finite alphabet as well as continuous sources.

In future work, we aim to  rigorously show the near-optimality
of our algorithm for continuous alphabet sources 
with the aid of the quantization scheme in~\cite[Section
    V.C]{SYLTAC2017POMDP}, which we already used, in a heuristic way, in
our simulations for a  continuous source.  Furthermore, we are
working on generalizing  the
results to the case where the channel is noisy and the
encoder has access to feedback from the channel; most results in our
paper seem to go through in this case too, with only  a slight modification of the
MDP formulation and the update equations.

We are also currently investigating a modified Q-learning approach
that uses a finite window of quantizers  and quantizer outputs as
the state of the code instead of  the current belief $\pi_t$.  The
benefits of such an approach include simpler implementation, fewer 
computations, and an explicit performance bound in terms of the
window length. The analysis for such codes becomes quite intricate
due to the fact that the notion of predictor stability
required for this method is stricter than that we used  in this paper,
motivating the study of alternative predictor stability conditions.

\appendices

\section{Supporting MDP Results}\label{appendix:A}
In the following, consider an MDP $(\mathsf{Z}, \mathsf{U}, P, c)$, where $\mathsf{Z}$ is the state space, $\mathsf{U}$ is the action space, $P(dz' | z, u)$ is the transition kernel, and $c : \mathsf{Z} \times \mathsf{U} \to \mathbb{R}_+$ is the cost function.
\begin{assumption}\label{assumption:ACOE}
    \begin{enumerate}
        \item[(i)] $c$ is continuous, nonnegative, and bounded.
        \item[(ii)] $\mathsf{Z}$ is a compact metric space.
        \item[(iii)] $\mathsf{U}$ is as a compact metric space.
        \item[(iv)] $P(\, \cdot\,  | z,u)$ is weakly continuous in $(z,u)$.
        \item[(v)] The family of functions $\{h_\beta : \mathsf{Z}\to
                \mathbb{R}, \; \beta \in (0,1)\}$, where $h_\beta(z)
                \coloneqq J^*_\beta(z) - J^*_\beta(z_0)$ for some fixed
            $z_0\in \mathsf{Z}$, is uniformly bounded  and equicontinuous.
    \end{enumerate}
\end{assumption}
The following is a standard result in the MDP literature, see
e.g.~\cite[Theorem 3.8]{Schal1993},~\cite[Theorem
    5.4.3]{HernandezLermaMCP} for related results.

\begin{lemma}\cite[Theorems 7.3.3 and 7.3.4]{yuksel2020control}\label{lemma:ACOE}
    Consider an MDP which satisfies  Assumption~\ref{assumption:ACOE}. Then there exist a constant
    $g^* \ge 0$ and a measurable function
    $f : \mathsf{Z} \to \mathsf{U}$ such that the stationary  policy
    $\gamma^* = \{f, f, \ldots\}$ is optimal for the average cost
    problem and $g^*$ is the optimal value function, i.e.,
    \[
        g^* = J^*(z) = J(z, \gamma^*) \text{\ \ for all $z\in
                \mathsf{Z}$.}
    \]
    Furthermore, $g^* = \lim_{\beta \uparrow
            1}(1-\beta)J^*_{\beta}(z)$. 
\end{lemma}

\begin{lemma}\label{lemma:ACOI}
    Let $g$ be a constant and $h : \mathsf{Z} \to \mathbb{R}$, $f : \mathsf{Z} \to \mathsf{U}$ be such that for all $z \in \mathsf{Z}$,
    \begin{equation}\label{eq:ACOI}
        g + h(z) \ge c(z,f(z)) + \int h(z')P(dz' | z, f(z))
    \end{equation}
    and
    \begin{equation}\label{eq:liminf}
        \liminf_{t \to \infty} \frac{1}{t} \mathbf{E}^\gamma_z \left[ h(Z_t)\right] \ge 0,
    \end{equation}
    where $\gamma = \{f, f, \ldots\}$ and $\{Z_t\}_{t\ge 0}$ is the
    state process under policy $\gamma$ and arbitrary given initial
    state $z\in \mathsf{Z}$. Then $g$ is an  upper bound
    to 
    the average cost of policy $\gamma$, i.e., 
    \begin{equation*}
        J(z, \gamma)\le g \text{ \ \ for all $z\in \mathsf{Z}$.}
    \end{equation*}
\end{lemma}
\begin{IEEEproof}
    By the law of iterated expectation, we have
    \begin{equation*}
        \mathbf{E}^\gamma_z \left[ \sum_{t=1}^T h(Z_t) \right] = \mathbf{E}^\gamma_z \left[ \sum_{t=1}^T \mathbf{E}^\gamma \left[ h(Z_t) | Z_{[0,t-1]}, U_{[0,t-1]}\right] \right].
    \end{equation*}
    By the Markov property and~\eqref{eq:ACOI},
    \begin{align*}
      \MoveEqLeft \mathbf{E}^\gamma \left[ h(Z_t) | Z_{[0,t-1]}, U_{[0,t-1]} \right] \\
        = & \int h(z')P(dz' | Z_{t-1}, U_{t-1}) \\
        = \; & \begin{multlined}[t]c(Z_{t-1}, f(Z_{t-1})) + \int h(z')P(dz' | Z_{t-1}, f(Z_{t-1})) \\ - c(Z_{t-1}, f(Z_{t-1}))    \end{multlined}           \\                                    
        \le \; & g + h(Z_{t-1}) - c(Z_{t-1}, f(Z_{t-1})).
    \end{align*}
                                                                           
    Substituting this into the previous equation, we obtain
    \begin{align*}
    \MoveEqLeft[5] \mathbf{E}^\gamma_z \left[ \sum_{t=1}^T h(Z_t) \right] \\
        \le \; & \mathbf{E}^\gamma_z \left[ \sum_{t=1}^T g + h(Z_{t-1}) - c(Z_{t-1}, f(Z_{t-1})) \right].
    \end{align*}
    Rearranging,
    \begin{equation*}
        \mathbf{E}^\gamma_z \left[ \sum_{t=1}^T c(Z_{t-1}, f(Z_{t-1}))\right]\le Tg + h(z) - \mathbf{E}^\gamma_z \left[ h(Z_T) \right].
    \end{equation*}
    Dividing by $T$ and taking the limsup,
    \begin{align*}
        J(z, \gamma) \le g + \limsup_{T \to \infty} \left( - \frac{1}{T}\mathbf{E}^\gamma_z \left[ h(z_T) \right] \right) \le g ,
    \end{align*}
    where we use~\eqref{eq:liminf} for the last inequality.
\end{IEEEproof}

The next result shows that if the discount factor $\beta$ is close enough to $1$,
a policy that is optimal or near-optimal for the discounted cost, is also
near-optimal in the average cost sense.

\begin{theorem}\label{theorem:discounted-near-optimal}
    Let Assumption~\ref{assumption:ACOE} hold. Then for every
    $\epsilon > 0$, there exists a $\beta \in (0,1)$ such that if
    a  stationary policy $\gamma_\beta=\{f_\beta,f_\beta,\ldots\}$
    satisfies  for some $\delta\ge 0$, 
    \begin{equation}
        \label{eq-delta}
        J_\beta(z, \gamma_\beta) \le J^*_\beta(z) +\delta 
    \end{equation}
    for  all $z\in \mathsf{Z}$,  then 
    \[
        J(z, \gamma_\beta) \le g^* + \epsilon   + (1-\beta)\delta 
    \]
    for  all $z\in \mathsf{Z}$,  where $g^*$ is the  optimal average  cost  (which is constant by Lemma~\ref{lemma:ACOE}) 
    
\end{theorem}

\begin{IEEEproof} As in Assumption~\ref{assumption:ACOE}, let
    $h_\beta(z) = J^*_\beta(z) - J^*_\beta(z_0)$ and define
    $ \hat{h}_\beta(z) \coloneqq J_\beta(z, \gamma_\beta) -
        J^*_\beta(z_0)$.  We will verify that the conditions of
    Lemma~\ref{lemma:ACOI} hold with $g =  g^* + \epsilon+(1-\beta)\delta$,
    $h = \hat{h}_\beta$, and $f = f_\beta$. Indeed, for any
    $\beta \in (0,1)$,
    \begin{align}
       \MoveEqLeft \hat{h}_\beta(z) + g^* - (g^* -
        (1-\beta)J^*_\beta(z_0)) \nonumber \\ & \qquad  + (1-\beta)\int \hat{h}_\beta(z')P(dz'
        | z,f_\beta(z)) \label{eq:beta} \\ 
         = \; & \hat{h}_\beta(z) + (1-\beta)J^*_\beta(z_0) \nonumber
      \\ & \qquad  +
        (1-\beta)\int \hat{h}_\beta(z')P(dz' | z,f_\beta(z))
        \nonumber
        \\ 
         = \; & \hat{h}_\beta(z) + (1-\beta)\int J_\beta(z', \gamma_\beta) P(dz' | z, f_\beta(z))                                                                       \nonumber \\
         = \; &
             c(z, f_\beta(z)) + \beta \int  J_\beta(z',
                \gamma_\beta)P(dz' | z,f_\beta(z)) - J^*_\beta(z_0)
                \nonumber \\
      & \qquad +
        (1-\beta)\int  J_\beta(z', \gamma_\beta) P(dz' | z, f_\beta(z))
          \nonumber                                                                                              \\
         & \; = c(z, f_\beta(z)) + \int \hat{h}_\beta(z')P(dz' | z,f_\beta(z)), \nonumber
    \end{align}
    where the third equality follows from the identity 
    \begin{equation*}
        J_\beta(z, \gamma_\beta) = c(z, f_\beta(z)) + \beta \int
        J_\beta(z', \gamma_\beta) P(dz' | z,f_\beta(z)).
    \end{equation*}
    Now consider the terms in \eqref{eq:beta}. By Lemma~\ref{lemma:ACOE}, there
    exists some $\beta_1 \in (0,1)$ such that
    \begin{equation}
        \label{eq:bound1}
        \left| g^* - (1-\beta)J^*_\beta(z_0) \right| \le \frac{\epsilon}{2}
        \text{ \ \  for all $\beta \in \left[\beta_1, 1\right)$.} 
    \end{equation}
    On the other hand,
    for any $\beta\in (0,1)$ choose $\gamma_\beta$ so that it is
    $\delta$-optimal, i.e., \eqref{eq-delta} holds. Then we have 
    \[ \| \hat{h}_\beta\|_\infty \le \sup_{z \in \mathsf{Z}}\left|
        J_\beta(z, \gamma_\beta) - J^*_\beta(z) \right| +
        \|h_\beta\|_\infty \le \delta  + \|h_\beta\|_\infty
    \]
    and therefore
    \[
        \left| (1-\beta)\int \hat{h}_\beta(z')P(dz' | z,f_\beta(z)) \right|
        \le  (1-\beta)( \delta  + \|h_\beta\|_\infty ).    
    \]
    By Assumption~\ref{assumption:ACOE}, $h_\beta$ is uniformly bounded
    over $\beta$, so there exists some $\beta_2 \in (0,1)$ such that for all $\beta \in \left[\beta_2, 1\right)$,
    \begin{equation}
        \label{eq:bound2}
        \left| (1-\beta)\int \hat{h}_\beta(z')P(dz' | z,f_\beta(z)) \right|
        \le  (1-\beta)\delta + \frac{\epsilon}{2}.
    \end{equation}
    Thus taking $\beta^* = \max\{\beta_1, \beta_2\}$ in
    \eqref{eq:bound1} and \eqref{eq:bound2},  we have for all
    $\beta \in \left[\beta^*, 1\right)$,
    \begin{align*}
        & g^* + \epsilon + (1-\beta)\delta  +\hat{h}_\beta(z) \\ \ge \; & c(z,
        f_\beta(z)) + \int \hat{h}_\beta(z')P(dz' | z,f_\beta(z)). 
    \end{align*}
    Finally,
    since $\hat{h}_\beta$ is bounded, \eqref{eq:liminf} is
    satisfied. Thus by Lemma~\ref{lemma:ACOI},
    \begin{equation*}
        J(z, \gamma_\beta) \le    g^* + \epsilon + (1-\beta)\delta.
    \end{equation*}
\end{IEEEproof}

\section{Algorithm 3}\label{appendix:B}
The following algorithm from \cite{Reznik} is used in our Algorithm~1 to quantize $\pi \in \mathcal{P}(\mathbb{X})$ to its nearest neighbor in $\hat{\pi} \in \mathcal{P}_n(\mathbb{X})$

\noindent \underline{\textbf{Algorithm 3: Predictor Quantization}} \cite[Algorithm 1]{Reznik}\label{algorithm:1}

\begin{algorithmic}[1]
    \Require $ n\ge1, \pi = (p_1, \ldots, p_m) $
    \For{$i=1$ \textbf{to} $m$}
    \State $k_i' = \lfloor np_i + \frac{1}{2} \rfloor$
    \EndFor
    \State $n' = \sum_i k_i'$
    \If{$n=n'$}
    \Return $(\frac{k_1'}{n}, \ldots, \frac{k_m'}{n})$
    \EndIf
    \For{$i=1$ \textbf{to} $m$}
    \State $ \delta_i = k_i' - np_i$
    \EndFor
    \State \textbf{Sort} $ \delta_i $ s.t. $ \delta_{i_1} \le \ldots \le \delta_{i_m} $
    \State $ \Delta = n'-n $
    \If{$ \Delta > 0 $}
    \State $k_{i_j} = \begin{cases}
            k_{i_j}' \quad   & j = 1,\ldots,m-\Delta    \\
            k_{i_j}'-1 \quad & j = m-\Delta+1,\ldots,m
        \end{cases}$
    \Else
    \State $k_{i_j} = \begin{cases} k_{i_j}'+1 \quad & j = 1,\ldots,|\Delta| \\ k_{i_j}' \quad   & j = |\Delta|+1,\ldots,m \end{cases}$
    \EndIf
    \State \textbf{return} $(\frac{k_1}{n}, \ldots, \frac{k_m}{n}) $
\end{algorithmic}


\begin{IEEEbiographynophoto}
{\bf Liam Cregg} received his B.A.Sc. degree in Mathematics and
Engineering from Queen’s University, Canada, and is currently completing his
M.A.Sc. degree in Mathematics and Engineering. He will be starting his Ph.D. degree
in Electrical Engineering at ETH Z{\"u}rich in 2024. His research interests
include stochastic control, reinforcement learning, information theory, and probability.
\end{IEEEbiographynophoto}

\begin{IEEEbiographynophoto}{\bf Tam\'{a}s Linder}
  (S'92-M'93-SM'00-F'13) received the M.S. degree in electrical
  engineering from the Technical University of Budapest, Hungary, in
  1988, and the Ph.D degree in electrical engineering from the
  Hungarian Academy of Sciences in 1992.  He was a post-doctoral
  researcher at the University of Hawaii in 1992 and a Visiting
  Fulbright Scholar at the Coordinated Science Laboratory, University
  of Illinois at Ur bana-Champaign during 1993-1994. From 1994 to 1998
  he was a faculty member in the Department of Computer Science and
  Information Theory at the Technical University of Bud apest. From
  1996 to 1998 he was also a visiting research scholar in the
  Department of Electrical and Computer Engineering, University of
  California, San Diego. In 1998 he joined Queen's University where he
  is now a Professor of Mathematics and Engineering in the Department
  of Mathematics and Statistics. His research interests include
  communications and information theory, source coding and vector
  quantization, machine learning, and statistical pattern recognition.
  Dr. Linder received the Premier's Research Excellence Award of the
  Province of Ontario in 2002 and the Chancellor's Research Award of
  Queen's University in 2003. He was an Associate Editor for Source
  Coding of the IEEE Transactions on Information Theory in
  2003-2004.\end{IEEEbiographynophoto}

\begin{IEEEbiographynophoto} {\bf Serdar Y\"{u}ksel} received his
  B.Sc. degree in Electrical and Electronics Engineering from Bilkent
  University; and M.S. and Ph.D. degrees in Electrical and Computer
  Engineering from the University of Illinois at Urbana-Champaign in
  2003 and 2006, respectively. He was a post-doctoral researcher at
  Yale University before joining the Department of Mathematics and
  Statistics at Queen’s University, Canada, where is now a
  Professor. His research interests are on stochastic control,
  information theory, and probability. Prof.\ Y\"{u}ksel is a
  co-author of three research books, a recipient of several awards,
  and has been an editor for a number of journals.
\end{IEEEbiographynophoto}

\end{document}